\documentclass[iop]{emulateapj}
\usepackage{xcolor}

\usepackage{natbib}
\begin{document}
\title{The Robotilter: An Automated Lens / CCD Alignment System for the Evryscope}

\author{Jeffrey~K.~Ratzloff\altaffilmark{1}, Nicholas~M.~Law\altaffilmark{1}, Henry~T.~Corbett\altaffilmark{1}, Octavi~Fors\altaffilmark{1,2}, Daniel~del~Ser\altaffilmark{1,2}}

\altaffiltext{1}{Department of Physics and Astronomy, University of North Carolina at Chapel Hill, Chapel Hill, NC 27599-3255, USA}
\altaffiltext{2}{Dept. de F{\'i}sica Qu{\`a}ntica i Astrof{\'i}sica, Institut de Ci{\`e}ncies del Cosmos (ICCUB), Universitat de Barcelona, IEEC-UB, Mart\'{\i} i Franqu{\`e}s 1, E08028 Barcelona, Spain}

\email[$\star$~E-mail:~]{jeff215@live.unc.edu}
%apj

%SPIE
%\title{The Robotilter: An Automated Lens / CCD Alignment System for the Evryscope}

%\author[a,*]{Jeffrey~K.~Ratzloff}
%\author[a]{Nicholas~M.~Law}
%\author[a]{Henry~T.~Corbett}
%\author[a,b]{Octavi~Fors}
%\author[a,b]{Daniel~del~Ser}

%\affil[a]{University of North Carolina at Chapel Hill, Department of Physics and Astronomy, 120 East Cameron, Chapel Hill, USA, 27599-3255}
%\affil[b]{Universitat de Barcelona, Dept. de F{\'i}sica Qu{\`a}ntica i Astrof{\'i}sica, Institut de Ci{\`e}ncies del Cosmos (ICCUB),  IEEC-UB, Mart\'{\i} i Franqu{\`e}s 1, E08028 Barcelona, Spain}

%\renewcommand{\cftdotsep}{\cftnodots}
%\cftpagenumbersoff{figure}
%\cftpagenumbersoff{table} 
%\begin{document} 
%\maketitle
%SPIE

%----------------------------------------------------------------------------------------
%	ABSTRACT
%----------------------------------------------------------------------------------------

\begin{abstract}

Camera lenses are increasingly used in wide-field astronomical surveys due to their high performance, wide field-of-view (FOV) unreachable from traditional telescope optics, and modest cost. The machining and assembly tolerances for commercially available optical systems cause a slight misalignment (tilt) between the lens and CCD, resulting in PSF degradation. We have built an automated alignment system (Robotilters) to solve this challenge, optimizing 4 degrees of freedom - 2 tilt axes, a separation axis (the distance between the CCD and lens), and the lens focus (the built-in focus of the lens by turning the lens focusing ring which moves the optical elements relative to one another) in a compact and low-cost package. The Robotilters remove tilt and optimize focus at the sub 10 $\mu m$ level, are completely automated, take $\approx$ 2 hours to run, and remain stable for multiple years once aligned. The Robotilters were built for the Evryscope telescope (a 780 MPix 22-camera array with an 8150 sq. deg. field of view and continuous 2-minute cadence) designed to detect short timescale events across extremely large sky areas simultaneously. Variance in quality across the image field, especially the corners and edges compared to the center, is a significant challenge in wide-field astronomical surveys like the Evryscope. The individual star PSFs (which typically extend only a few pixels) are highly susceptible to slight increases in optical aberrations in this situation. The Robotilter solution resulted in a limiting magnitude improvement of .5 mag in the center of the image and 1.0 mag in the corners for typical Evryscope cameras, with less distorted and smaller PSFs (half the extent in the corners and edges in many cases). In this paper we describe the Robotilter mechanical and software design, camera alignment results, long term stability, and image improvement. The potential for general use in wide-field astronomical surveys is also explored. 

\end{abstract}

%SPIE
%\keywords{Lens / Camera alignment, optical alignment, Telescope alignment}

% Include email contact information for corresponding author
%{\noindent \footnotesize\textbf{*}Jeffrey~K.~Ratzloff,  \linkable{jeff215@live.unc.edu} }

%\begin{spacing}{2}   % use double spacing for rest of manuscript
%SPIE

%----------------------------------------------------------------------------------------
%	INTRO
%----------------------------------------------------------------------------------------

\section{Introduction}

Commercial camera lenses are used on SuperWASP \citep{2006PASP..118.1407P}, HAT \citep{2004PASP..116..266B}, HatNet and HATSouth \citep{2018haex.bookE.111B}, KELT \citep{2007PASP..119..923P}, XO \citep{2005PASP..117..783M}, MASCARA \citep{2017A&A...601A..11T}, and other transiting exoplanet surveys to reach as much as 1000 square degree fields of view. Other surveys types such as the ASAS-SN (supernova) \citep{Shappee2014}, Pi of the Ski (gamma ray bursts) \citep{2013A&A...551A.119P}, and Fly's Eye (asteroid detection) \citep{2013ASPC..475..369C} also use camera lenses to reach wide sky coverage. The Evryscope (described in detail in \citealt{2019arXiv190411991R}) also uses camera lenses to provide continuous all-sky coverage with fast cadence, aimed at finding rare short-time events. Each of these surveys pair the camera lenses with compact CCD cameras to achieve the FOV and pixel sampling necessary at a modest cost. They have discovered a variety of photometrically variable objects including exoplanets, binaries, stellar phenomenon, and galactic events.

These types of wide field surveys and many others including the Evryscope are susceptible to image quality challenges from CCD / lens misalignment (tilt) and sub-optimal focus. The tilt and focus challenges are driven by two primary factors - mechanical and software. The mechanical challenge is to align the optics with respect to the camera to the level necessary to minimize PSF differences across the image; for very wide fields and fast optics, this requires precision beyond typical machining and assembly tolerances. The software challenge involves optimizing 4 degrees of freedom (2 in tilt, 1 in lens / CCD separation, and 1 in focus position - see Figure \ref{fig:robotilter_concept} later in the manuscript) with severe degeneracies and local extremes. The software solution also requires a method to measure image quality, across all regions, in the presence of tilt -- a non trivial task. The image quality measurement must be capable of handling PSF differences due to focus, off-axis aberrations, and SNR variations.

There are very few discussions concerning removing image tilt or optimising the focal plane in wide-field astronomical surveys, and we found none that use an integrated tilt removal system as part of their main instrument design. Conversely, the majority of the surveys \citep{2017A&A...601A..11T, 2007PASP..119..923P, 2006PASP..118.1407P, 2004PASP..116..266B} discuss the challenges of PSF distortions and focal plane issues from the wide fields. A common struggle is the negative impact on the photometric precision, poor performance on dim stars, and the difficulty in reaching the sub-percent level required for typical exoplanet searches. Extensive software development is put into the calibrations, pipeline, aperture photometry, and systematics removal of each of these instruments to try and maximize light curve quality given the challenges of a very wide field. Several new solutions resulted from these struggles including multi-aperture forced photometry, wide-field astrometry solutions, and methods for maximizing under-sampled PSFs. A reliable method to remove image tilt and optimize the focal plane would complement and reduce the burden on the software solution, and potentially improve limiting magnitude and light curve precision. With the Evryscope F1.4 optics and 384 sq.deg. individual camera FOV (among the most aggressive of the current wide field surveys), image tilt removal is more necessity than option.

Photometric surveys that use small consumer telescopes or custom optics instead of camera lenses, for example MEarth \citep{2008PASP..120..317N} and PROMPT \citep{2005NCimC..28..767R} and many others, typically have FOVs on the order of a few sq. deg. or less. The smaller FOVs tend to have much slower optics than the camera lens based surveys, combined with finer pixels, lessens the PSF challenges due to optics misalignment. These surveys function well without the need of an advanced tilt removal solution.

Turning to larger aperture instruments with wide fields, the process typically starts by fixing the primary mirror and aligning the secondary, then progressing to any other elements, and finishing with the CCD. Pan-STARRS \citep{2010SPIE.7733E..0EK} used an auto-reflecting telescope, developed a custom alignment software, combined with $\approx$ 1 month of observing time to realign the optics to a level of tens of microns. LSST \citep{2012PASP..124..380S} will align the optics and CCD using laser targets fixed to the primary and the off-axis aberration characteristics of the wide field to characterize and remove tilt. They plan to also align sequentially proceeding from the secondary to the CCD, and simplify the procedure by designing the primary and tertiary from the same blank (so that they are fixed in alignment). ZTF \citep{2018arXiv180210218B, 2018SPIE10702E..4KD} uses on sky images to match each of the 16 CCD portions to the focal plane within $\approx$ 10$\mu m$, in order to meet their photometric precision requirements. Although these are considerably different instruments on entirely different cost and complexity scales, some of the alignment principles used helped confirm our solution ideas for the camera lens / CCD based Evryscope. Moving only one element and holding all others fixed simplifies the process, and in many cases is the only practical way to avoid alignment degeneracies. Using on-sky images offers the advantage of the same focus position and conditions as science images. Reliably measuring image quality across wide fields with significant PSF distortion is challenging, even more so in the condition of under-sampled PSFs. The instrument differences are also evident, the Robotilters must be simple, economical, and avoid complex components such as laser targets and resource intensive steps such as camera disassembly / shimming / reassembly for alignment. We also note that moving individual lens elements relative to each other is not an option as the lenses are sealed, and there are no external adjustments for the lens components. The Robotilter solution must be modular to work on all cameras and fields, and ideally could be scaled to work on other instruments. 

The primary source of the misalignment stems from the way wide field instruments using camera lens attach the lens to the CCD through a series of elements. Typically, the CCD is mounted to a camera housing, the housing is mated to a filter wheel, which is in turn mated to a lens mount (bayonet ring), and the lens turns and locks onto the bayonet ring. The manufacturing tolerances, and the multiple mating points and assembly steps causes a misalignment in the lens and CCD. Even a slight tilt will result in an unacceptable increase in size of the PSF FWHM towards the edges and corners of the image.

To estimate the increase in PSF size due a tilted image, we estimate the blur diameter $B_{d} \propto \frac{\Delta}{F}$, where $\Delta$ is the defocus and F is the F number of the lens. The defocus is dependent on the tilt ($\theta$) between the lens and CCD and on the distance (\textit{d}) of the source from the center of the CCD. The estimated PSF increase (in pixels) is then $PSF_{inc} \propto \frac{\theta \times d}{F \times pixel size}$. For sources near the edges and corners of the field, the effect is strong given the aggressive F1.4 optics of the Evryscope. We originally estimated that for the Evryscope cameras, even a very small tilt at the level of a 5 $\mu m$ difference in \textit{opposite edges} of the CCD (a .02 degree tilt) would result in excessive ($\approx$ one pixel) PSF increases toward the edges of the CCD.

The lens mounting surface (the region where the lens contacts the filter wheel and subsequently where the filter wheel contacts the CCD housing) is 3 times larger than the CCD. This relaxes the 5 $\mu m$ tilt difference at the CCD edges to $\approx$15 $\mu m$ difference in \textit{opposite edges} of the lens mounting surface. This is still challenging given the Evryscope fast optics. For example, a .100 mm thickness difference (from normal machining tolerances) in opposing edges of the lens mounting ring results in a .4 degree tilt between the CCD plane and the lens focal plane. The PSF FWHM could increase by as much as double or triple in this situation. In images taken with misaligned optics (hereafter misaligned images), the PSF shape is often compromised leading to elongation - in severe cases the width of one axis might grow to double or triple the width of the long axis of the elongated PSF. In most cases the center of the image is well focused, and a severely tilted image will have an edge-of-CCD region that is out of focus below the focal plane and a region opposite that is out of focus above the focal plane. The elongation and distortion of these two regions are different. We interpret these observed effects as out of focus regions in the presence of field aberrations (likely dominated by coma and astigmatism.)

Consumer lenses are designed to operate over a wide focus range, however finding the lens focus position that focuses each region in the image similarly well is a challenge. The steep light cone associated with fast lenses increases the demands on optical design, manufacturing precision, and material quality to achieve required performance levels. If severe enough, the tilt and focus problems will compromise the desired science goals by causing errors in astrometry, aperture photometry, inconsistent star observations, and increased noise in the light curves. 

In this work we describe several new mechanical design and software solution approaches, develop a novel mounting design, and combine them into a compact and effective tilt removal and focus optimization system. The Robotilters are an inexpensive (they cost only a few percent of the total Evryscope instrument cost), completely robotic, on-sky tilt removal system for the very wide field Evryscope cameras. We demonstrated the concept and showed initial results in \cite{2016SPIE.9908E..0WR}, here we describe the full solution and results. The Robotilters take 2 hours to run, remove tilt to the sub 10 $\mu m$ level (as measured from opposite sides of the lens mounting surface or equivalently the Robotilter servo shafts) on a typical camera, optimize the focus across the image, and remain stable for multiple years once the final solution is found. We show the Evryscope image quality challenges introduced by tilt and focus, and their detrimental effect on limiting magnitude, astrometry, PSFs, and SNR. We demonstrate our solution to remove tilt and optimize focus across the image. We also briefly discuss the potential of the Robotilter design for use on other wide field surveys. We installed the Robotilters in November 2015 and began testing hardware and camera alignment software on select cameras in early 2016. All cameras were aligned by mid 2016 and have been stable for three years with only minor focus adjustments. The Evryscope hardware and optics, combined with the moderate night-to-night temperature changes at the CTIO observing site do not require constant refocusing; instead only periodic refocusing is done in response to seasonal temperature swings. 

We discuss the system requirements in \S~\ref{section_constraints}, and show the Robotilter system in \S~\ref{section_robos}. The optimization software is explained in \S~\ref{section_software}, and the alignment results and image improvement presented in \S~\ref{section_combined_results}. We discuss the results in \S~\ref{section_discussion}, and conclude in \S~\ref{section_summary}.

%----------------------------------------------------------------------------------------
%	CONSTRAINTS
%----------------------------------------------------------------------------------------

\section{System Requirements} \label{section_constraints}

\subsection{Science Requirements}

\subsubsection{Image Quality Requirements}

The planned Evryscope surveys \citep{2015PASP..127..234L} requires sub percent level photometric precision on stars $m_{g} = 12$ and brighter and few percent level on stars 12 $< m_{g} <$ 15, continuously in each 2 minute exposure. In order to achieve this level of photometric precision, our models show the PSF FWHM needs to be between two and four pixels to avoid over or under-sampling, the loss of significant signal to background, or the necessity for large photometric apertures (the circle used to define the pixels included in the PSF, hereafter photometric aperture). The limiting magnitude and photometric precision benefits from the PSFs being round without distortion, and they need to be consistent across the image.

The Evryscope pixel scale was driven by several requirements, mostly by the very wide field-of-view ($\sim$ 10,000 sq. deg.), the signal-to-noise (SNR) required to detect transits, the limiting magnitude required to achieve enough sources, and the target of less than 30 cameras (to limit overall instrument complexity), along with using commercial components (reliability and cost). Given the final selection of components, the pixel scale is 13 arcsec per pixel. This is acceptable, with the main concern being that too coarse of a pixel scale results in a higher fraction of blended sources (nearby stars being blended in the pixel).

\subsection{Functional Requirements}

\subsubsection{Mechanical Constraints - The Evryscope Telescope Modules}

The Evryscope \citep{2019arXiv190411991R} is an array of 27 identical individual telescopes mounted into a hemispherical shell, called the mushroom, with a single common telescope mount. It uses Rokinon 61mm effective diameter F1.4 lenses paired to 28.8 MPix KAI29050 CCDs. The Evryscope uses FLI CFW-5-1 filter wheels, which have the capacity to accommodate 5 different filters. We use a single science filter (a modified SDSS G) with the other positions populated with sunshields used to protect the system in the event of a dome failure during the day. The science filter is designed with the parallelism and surface quality necessary to avoid measurable aberrations given the Evryscope lens and CCD specifications. All images shown and referred to in this manuscript were taken in the single science filter mode.

\begin{figure}[tbp]
\centering
%apj
\includegraphics[width=1.0\columnwidth]{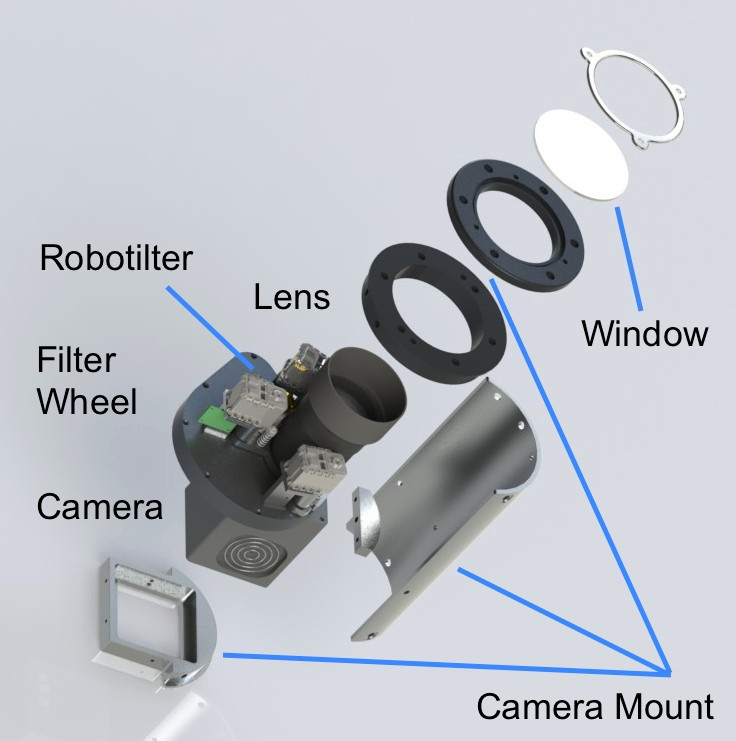}
%SPIE
%\includegraphics[width=3.5in]{Camera_assembly_exploded_wrobo_crop.png}
\caption{The Evryscope telescope modules, showing the mount, CCD camera, filter wheel, lens, optical window, and the Robotilter automated alignment system. The Robotilter uses three precision servos to adjust the separation and rotation between the lens and CCD to remove tilt and align the optical system. A separate servo is used to adjust the lens focus.}
\label{fig:camera_mount}
\end{figure}

Camera mounts support and point the telescopes to form modules as shown in Figure \ref{fig:camera_mount}. The telescope modules have to be as compact as possible to keep the size of the mushroom to less than 6 feet in diameter to meet the size and weight constraints of the Evryscope and the CTIO observing dome. The Robotilters need to fit into a small 8"x6"x4" space of the telescope module mounting on top of the filter wheel, between the lens and CCD. The Evryscope budget and resource limitations require the Robotilters to be simple with minimal components and few moving parts, without exotic materials, and using only readily available hardware. The unit cost target is \$1000 or less, assembly time must be modest, and they need to perform reliably without human intervention.

\subsubsection{Operations Constraints - Automation}

The Evryscope currently has 22 cameras with the capacity for 5 more and is located at the remote observing site in CTIO, Chile. The system operates robotically, averaging 5000 images per night with $\approx$300,000 sources per image. Using the conventional method to fix the slight misalignment between the optics and CCD (by inserting shims or small thumb screws between the CCD and the lens and manually adjusting the thickness in an iterative way) is unfeasible for a system like the Evryscope. The time and resource requirements to adjust the very small physical distances necessary to correct PSFs are excessive. We demonstrate in \S~\ref{section_effects} that leaving the lens / CCD misalignment uncorrected has a negative effect on light curve precision and reduces detection efficiency. This requires our tilt correction solution be automated, efficient, and repeatable, and remain consistent for multiple years once aligned. In addition to tilt removal, a focus step must optimize quality across the image with the ability to automatically compensate for temperature changes.

\subsubsection{Image Quality Measurement} \label{sub_section_image_quality_maeasurements}

\begin{figure}[tbp]
\centering
%apj
%\includegraphics[width=1.0\columnwidth]{combined_robo_5.jpg}
%\includegraphics[width=1.0\columnwidth]{ML0094214_20150725_021554_fwhm_plot.png}
\includegraphics[width=1.0\columnwidth]{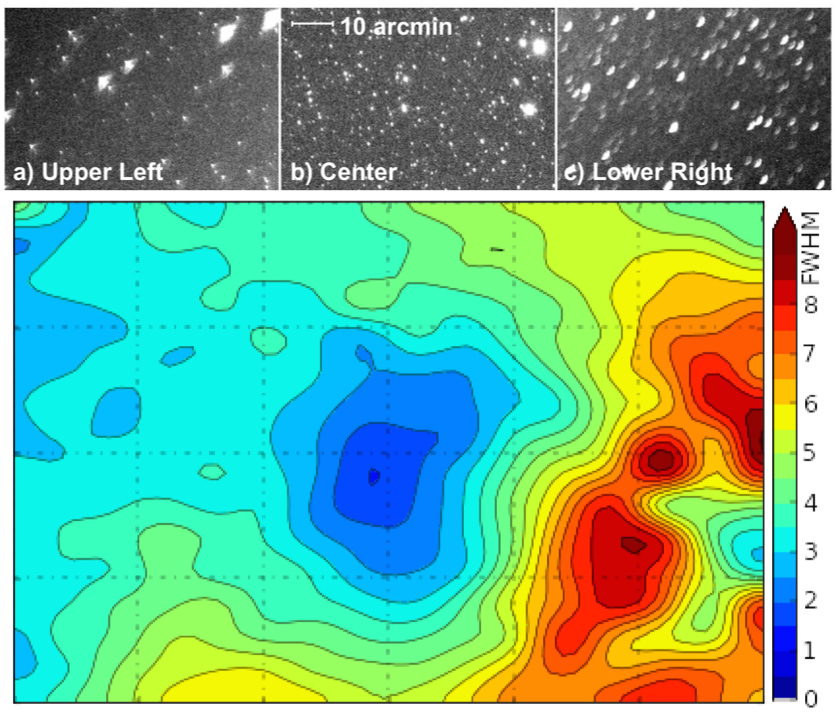}
%SPIE
%\includegraphics[width=3.5in]{combined_robo_5_all.png}
\caption{\textit{Top: a)} An initial deployment (pre-Robotilter) image from the polar facing camera showing a 300 x 200 pixel closeup of problematic upper left corner. \textit{b)} Closeup of the center of the same image. \textit{c)} Closeup of the problematic lower right corner of the same image. \textit{Bottom:} A PSF FWHM contour plot for the full image, demonstrating the challenge in quantifying PSF quality in severely tilted images and the lack of distinction for regions out of focus on opposite sides of the focal plane.}
\label{fig:image_quality_field}
\end{figure}

A small physical difference within normal machining tolerances of only $\approx$75 microns (a few thousandths of an inch as it is commonly expressed in CNC machining precision) can significantly degrade PSF quality in a wide-field image, especially with fast optics. Figure \ref{fig:image_quality_field} shows a pre-Robotilter image with tilt from the upper left corner to the lower right corner. The image center is well focused, while the upper left and lower right corners are out of focus and on opposite sides of the focal plane. The opposing corner regions show distinct differences in PSF shape, distortion, and extent. A PSF FWHM contour plot is also shown for the same image, demonstrating the challenge in quantifying PSF quality in severely tilted images. The low quality of the lower right corner is well captured by the high FWHM values, and the high quality of the center is well captured by the low FWHM values. However, the low quality of the upper left corner is not well captured by the low FWHM values nor is there a distinction for regions out of focus on opposite sides of the focal plane. This turned out to be the most difficult challenge of the Robotilter project. We developed our own image quality metric and image comparison method as explained in \S~\ref{section_software} in order to remove image tilt.

We use Source Extractor \citep{1996A&AS..117..393B} for all measurements taken to determine image quality, regardless of the quality metric used (FWHM, Strehl, number of sources, or custom quality metrics described later in the manuscript). Each image used in the Robotilter algorithm is first processed with the standard Evryscope pipeline calibrations and image quality checks. Master-flats and darks are applied depending on the camera the image was taken with, bad pixels are masked, and images with very high overall background levels or obvious quality issues (clouds, streaks, or jitters) are discarded. For further details of the Evryscope pipeline and data processing we refer the reader to our Evryscope instrument paper \cite{2019arXiv190411991R}. Source Extractor uses a threshold above a local average of flux values (measured at each pixel) to detect sources above this level. Sources with adjacent pixels above this local average are counted as detections. A centroiding step provides the source location and a photometric aperture (a circle encompassing a radius of pixels) is used to sum the flux from the source.

Source Extractor offers a variety of input settings, including background significance level, minimum number of pixels in the aperture, and aperture size (expressed as a pixel radius). As a starting point, we relied on the settings from our photometric pipeline. Given the FOV and pixel scale of the Evryscope (13 arcsec / pixel) we expect and find that most use-able sources are 3$\sigma$ above the background, the source PSFs have pixel counts ranging from a few for dim sources and $\approx$100 for the very brightest non-saturated sources, and the average best photometric aperture is modest in size at $\approx$3 pixel radius. Again, we refer the reader to \cite{2019arXiv190411991R} for further Evryscope instrument details. For the Robotilter algorithm we require a 4$\sigma$ above background threshold and a minimum pixel count of 15 per PSF; the more stringent requirements filter very poorly sampled and dim sources unlikely to be use-able in calculating PSF quality.

For the sources with only a few pixels in the PSF, we include these in a number of sources count. The count is used as a separate quality metric, found to be independent of others such as the FWHM and Strehl. The sources count is very susceptible to observing conditions and to the observation field. However as we describe later in the manuscript, for the observations taken to align an individual camera we hold the field constant and take the observations over a short range of time with similar sky conditions to minimize the bias. Although we do not directly use the dimmer sources in the standard quality metrics, they make a valuable separate contribution in determining detected sources.

We tested a small range of input settings near the values described in the previous paragraphs (expected to be reasonable for given the Evryscope instrument characteristics) and found they did not help the FWHM or Strehl performance. As we show throughout this work, the the FWHM especially struggles to perform over the likely range of image misalignment. This is a result of the metric, the coarse pixel sampling, the wide-field and amount of tilt, and not due to software settings. The Source Extractor FWHM value is determined by fitting a two-dimensional Gaussian to the extracted PSF and calculating the weighted average of the width at half the maximum value. As a second check we used our own photometric aperture to extract pixel counts and values (thus only relying on Source Extractor for the source locations) and recalculated the FWHM directly and found no noticeable difference.

The testing and analysis of the more traditional quality metrics like the FWHM and Strehl ultimately influenced the approach we took to finding a solution the image quality challenge. While these traditional quality metrics struggle in many image regions and tilted images, they can work in limited ranges and if the PSF pixel sampling is good enough. If the metrics are somewhat independent (they succeed or fail in different situations), they might still be leveraged together to form an effective quality metric. Later in this work we show that we combined several traditional metrics and some custom ones as components to form our final composite type quality metric. This approach produced a reliable, simple, and fast solution for the Evryscope images.

\subsubsection{Optimal Image Focus}

\begin{figure}[tbp]
\centering
%apj
%\includegraphics[width=1.0\columnwidth]{combined_robo_6.jpg}
%\includegraphics[width=1.0\columnwidth]{ML0453714_20150725_021529_fwhm_plot.png}
\includegraphics[width=1.0\columnwidth]{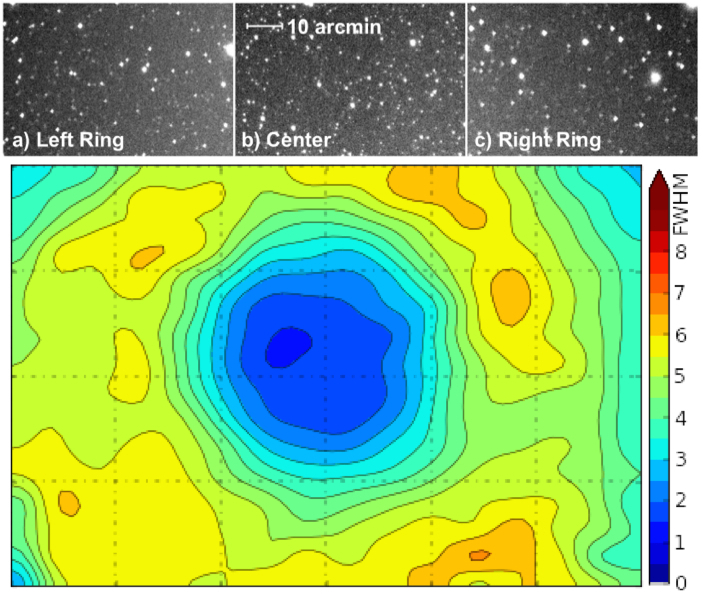}
%SPIE
%\includegraphics[width=3.5in]{combined_robo_6II_all.png}
\caption{\textit{Top:} An image from a tilt corrected zenith facing camera but without the Robotilter focus optimization. This image solution is found by maximizing the center image focus at the expense of the outer regions. \textit{ a)} Shown is a 300 x 200 pixel closeup of the left side of the image showing the problematic defocused ring. \textit{ b)} Closeup of the well focused center of the same image. \textit{ c)} Closeup of the right side of the image showing the same problematic ring. \textit{Bottom:} A PSF FWHM contour plot for the full image, demonstrating the challenge in optimizing the focus of the entire image - here resulting in unnecessarily large PSFs toward the outer field and a ring-like feature.}
\label{fig:image_quality_focus}
\end{figure}

An image taken from a camera with well aligned optics (hereafter a well aligned image) can be brought into focus by measuring the average PSF FWHM or Strehl for a small region in the center of an image. This on-axis focusing prescription works well in almost all situations, however in a very wide-field image like the Evryscope a focus optimization can provide an improvement in average PSF quality across the image. Figure \ref{fig:image_quality_focus} shows a flat image with no tilt and a very well focused image center, but with a compromise in PSF quality as the radial distance increases resulting in a ring feature. In \S~\ref{section_software} we discuss our solution to defocus the image center in the direction and amount that optimises the overall image focus. We were able to incorporate this step with the tilt removal procedure so that our final solution concurrently removes image tilt and optimizes image focus.

\subsubsection{Wide Field Survey Issues}

Wide field surveys suffer from additional challenges that degrade image quality including field aberrations (predominantly coma, astigmatism, and curvature) and spherical aberrations (SA). The field aberrations and SA challenges are due to the difficulty of achieving a very wide field of view with a large aperture lens. The effects can be mitigated (but not completely removed) by the lens and CCD choice, FOV requirements, proper image calibration, photometric aperture selection, and removal of systematics.

An additional significant challenge for the Evryscope survey is lens vignetting. Although this issue is present in most lenses, it is normally more impact-full in wide field surveys. The vignetting is assumed to be radially symmetric and centered in the image, however the properties and magnitude must be characterized with photometric calibrations and are unique to each camera assembly.

Pixel drift (drift) is a challenge all telescopes face, even more so for a wide field like the Evryscope. The drift primarily arises from the telescope misalignment or camera flexure. The Evryscope is aligned based on the polar facing camera as all tracking is from this pointing. The higher elevation cameras are over 90$^{\circ}$ in declination from the polar camera, so even a small misalignment can be challenging for the Evryscope. 

In this work, we do not address lens choices or how they might affect field aberrations, lens vignetting, and SA; nor do we explore telescope mount designs to prevent pixel drift. The Coma, lens vignetting, SA, and drift challenges as they relate to the Evryscope are described in \cite{2019arXiv190411991R} and \cite{2016SPIE.9908E..0WR}, in this work we concentrate on the image quality challenges introduced by tilt and focus and our solution to remove tilt and optimize focus across the image.

To summarize the ideas in this section, a corrected Evryscope image will be flat enough that the PSFs in opposing corners or sides will be similar in shape and size (within a pixel difference in extent from the center of the PSF). This maximizes the dim source detection (limiting magnitude) and avoids large photometric apertures which can degrade light curve precision. The corrected images will still have differences in PSFs in the image center versus the corners and the corners especially will still have elongations due to the effect of aberrations inherent to the optical system.

%----------------------------------------------------------------------------------------
%	ROBOTILTERS
%----------------------------------------------------------------------------------------

\section{THE ROBOTILTER DESIGN} \label{section_robos}

\begin{figure}[tbp]
\centering
%apj
%\includegraphics[width=1.0\columnwidth]{ccd_lens_tilt_anotated_cropped.png}
\includegraphics[width=1.0\columnwidth]{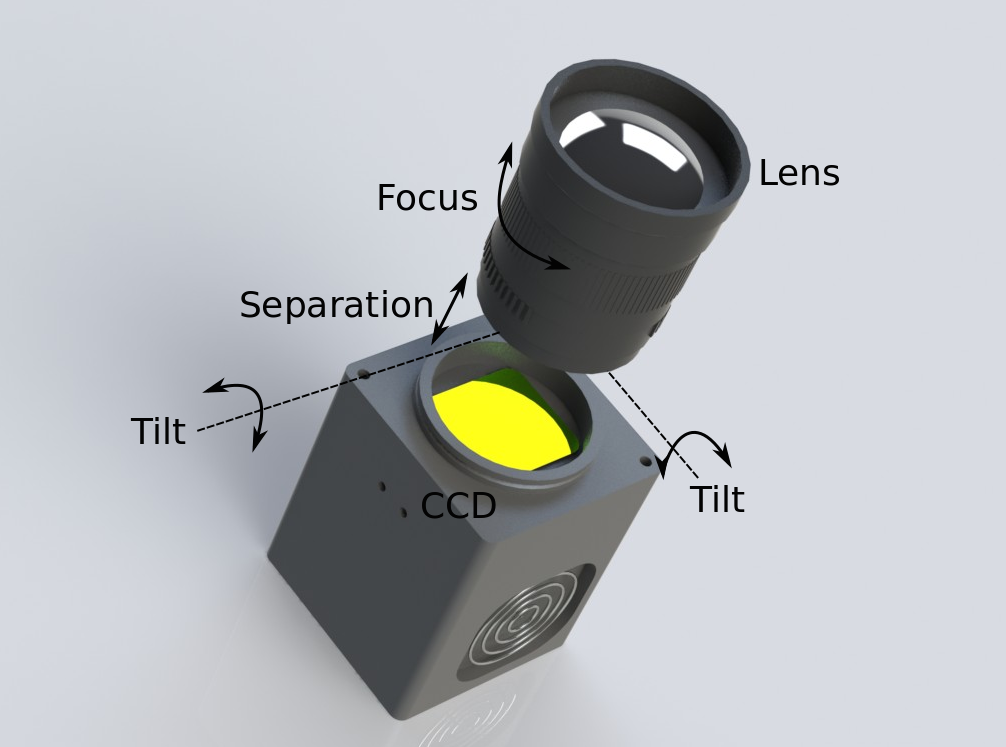}
%SPIE
%\includegraphics[width=3.5in]{ccd_lens_tilt_anotated_cropped.png}
\caption{The Robotilter concept: the lens is moved relative to the CCD to remove tilt and optimize image quality.}
\label{fig:robotilter_concept}
\end{figure}

The Robotilter approach is to move the lens relative to the CCD, and adjust the focus via the lens focus and separation distance. Figure \ref{fig:robotilter_concept} shows the arrangement, resulting in 4 degrees of freedom - 2 tilt axes (also known as tip/tilt, referred to as tilt throughout this manuscript), a separation axis (the distance between the CCD and lens), and the lens focus (the built-in focus of the lens by moving the lens barrel which moves the optical elements relative to each other). We elected to fix the CCD camera and move the lens because the lens is the smaller and lighter of the two components. 

\begin{figure*}[ht]
\centering
%apj
%\includegraphics[width=2.0\columnwidth]{IMG_0686.JPG}
\includegraphics[width=2.0\columnwidth]{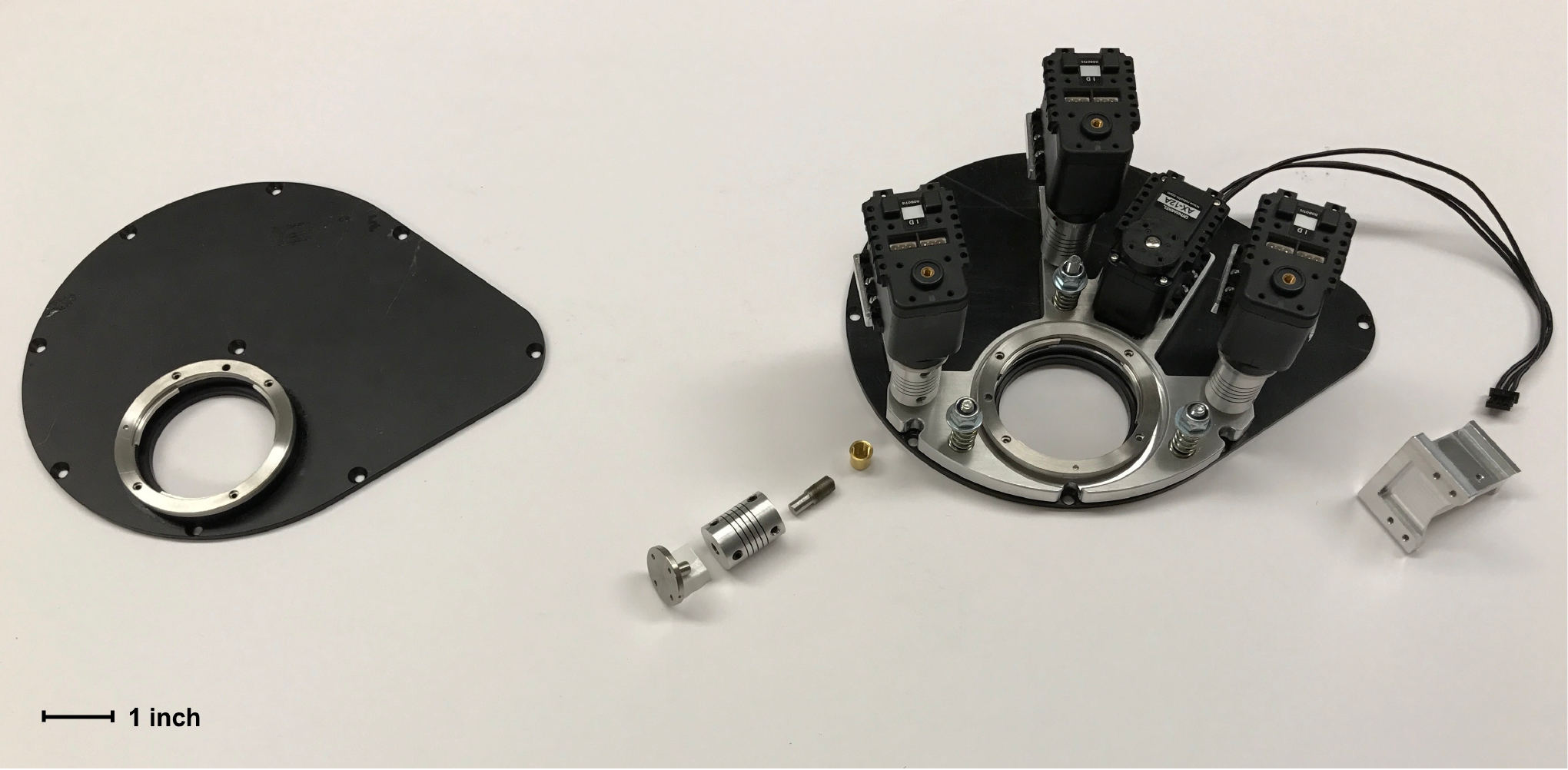}
%SPIE
%\includegraphics[width=1.0\columnwidth]{IMG_0686.pdf}
\caption{\textit{Left:} The conventional mounting system with the lens bayonet ring fixed to the top of the filter wheel. \textit{Right:} The Robotilter design. The lens bayonet ring is instead fixed to the lens base-plate. Three threaded shafts suspend the base-plate above the filter wheel top and are turned by precision servos. As each shaft turns, the base-plate moves up or down at the shaft axis relative to the filter wheel top and adjust the tilt and separation of the base-plate and lens relative to the CCD. A fourth servo is attached to a brass gear which contacts a plastic gear track fixed to the lens; as the brass gear turns, the lens focus adjusts. The four degrees of freedom - the two tilt axis, the separation axis, and the lens focus can be optimized.}
\label{fig:robotilter_design}
\end{figure*}

\begin{figure*}[ht]
\centering
%apj
%\includegraphics[width=2.0\columnwidth]{robotilter_exploded_anotated.png}
\includegraphics[width=2.0\columnwidth]{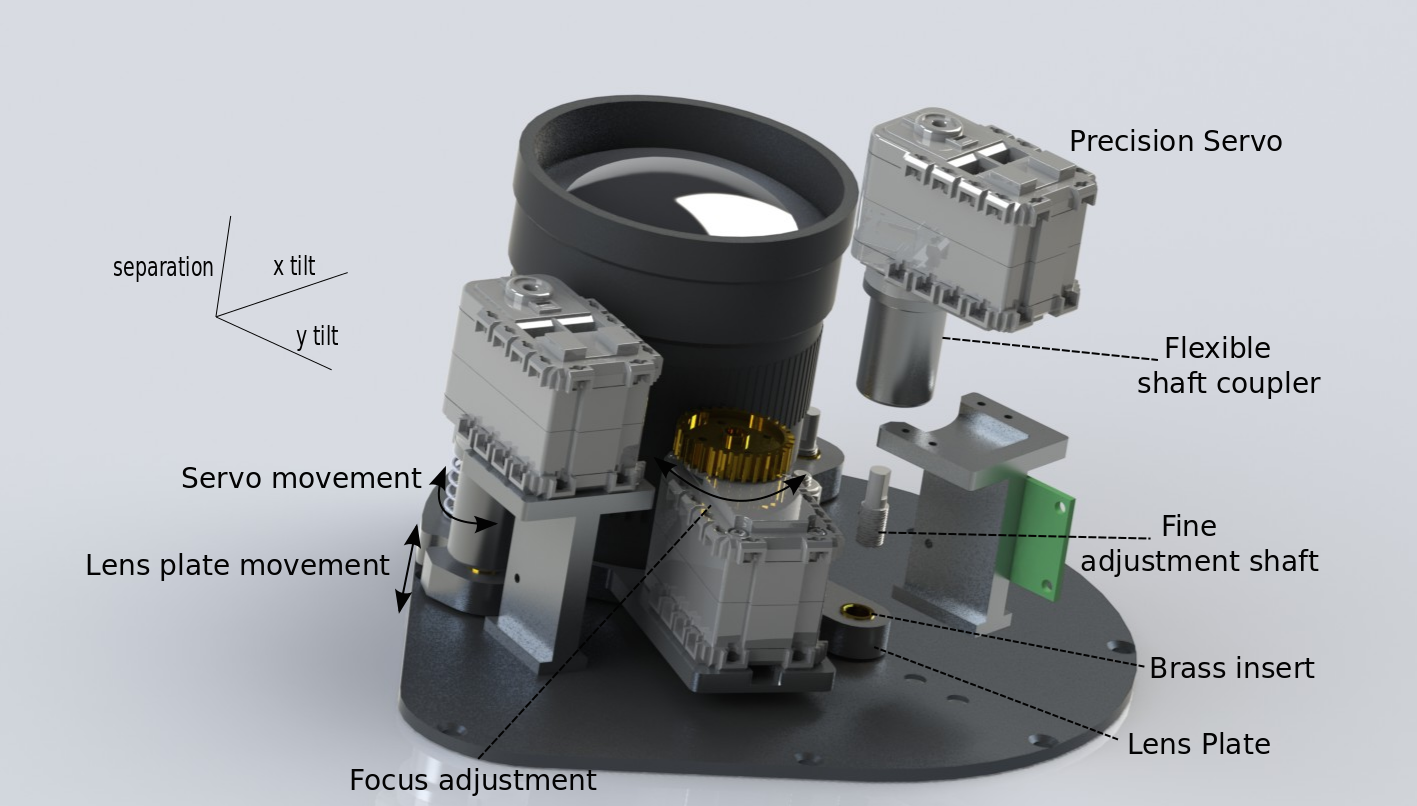}
%SPIE
%\includegraphics[width=1.0\columnwidth]{robotilter_exploded_anotated.png}
\caption{The Robotilter automated tilt removal and focus optimization mechanism. Servo movement adjusts the lens plate relative to the CCD. Exploded views of the servo, flexible shaft coupler (to prevent binding), fine adjustment shaft, and brass insert are shown along with the focus adjustment servo and gear.}
\label{fig:robotilter_anotated}
\end{figure*}

A conventional mounting system fixes the lens to the CCD by mounting the lens bayonet ring to the filter wheel top (which is in turn fixed to the CCD housing)\citep{2006PASP..118.1407P, 2004PASP..116..266B, 2018haex.bookE.111B, 2007PASP..119..923P, 2017A&A...601A..11T, Shappee2014}. The lens turns onto the bayonet ring and locks into place via a spring and set screw\footnote{https://www.canon.ie/lenses/tech-guide/}. The Robotilter instead replaces the fixed mounting system with a movable lens base-plate as shown in Figures \ref{fig:robotilter_design} and \ref{fig:robotilter_anotated}. The lens bayonet ring now fixes to the lens base-plate and the base-plate is suspended above the filter wheel top by 3 threaded shafts. As each shaft turns, the base-plate moves up or down at the shaft axis relative to the filter wheel top. The three shafts are positioned in a triangular pattern and are held firmly against the filter wheel top by tension springs regardless of telescope orientation. The stainless steel shafts use a very fine 80 threads per inch (TPI) for precise movement capability and the lens base-plate has pressed-in brass inserts for smooth operation. Each of the threaded shafts are connected to a flexible coupler which fixes the input and output in rotation, but allows for a small angular difference and for changes in length. This prevents binding as the base-plate is moved. Servo piers are attached to the filter wheel top providing a secure mounting point for the servos. 

Dynamixel MX-12 servos turn the couplers and shafts. Combined movements of the servos adjust the tilt of the base-plate and lens relative to the CCD. The lens can also move toward or away from the CCD without changing the tilt if the three servos are moved in the same direction and in equal steps. To adjust the lens focus a fourth servo is attached to a brass gear which contacts a plastic gear track fixed to the lens; as the brass gear turns, the lens focus adjusts. In this way, the four degrees of freedom (the two tilt axis, the separation axis, and the lens focus) can be optimized.

\subsection{Mechanical Design Features}

The Robotilter is designed for precise movements of the lens relative to the CCD. The tilt adjustment servos are controllable to within 2 degree accuracy in rotation and when coupled to the 80 TPI adjuster, the tilt and lens / CCD separation can be adjusted in increments (theoretically) as fine as .0001 inch (3 $\mu m$). The fine movements are consistent and repeatable, and at the sub 10 $\mu m$ precision necessary to remove tilt and optimize focus. The servos can be turned multiple rotations, and the lens base-plate has enough travel ($\pm$ 15,000 steps) to cover the range necessary ($\pm$ 6000 steps) to find the optimal position. Once the final solution is found, the servos can be locked and remain in place reliably.

The optical path is sealed using several approaches to prevent light loss or dust contamination. Critical mating surfaces are recessed, the lens bayonet ring and the lens base-plate for example, to produce an overlap. At the movement interface, a light-trapping ring extends below the lens base-plate to prevent stray light from entering the optical system without impeding lens movement. Light and dust trapping foam is used between the base-plate and the filter wheel top as an additional seal.  

The servo piers are slotted to match the bottom surface of the servos, and the filter wheel top is slotted to match the bottom of the servo piers. The shaft couplers use dual setscrews to securely fasten to the shafts, and each shaft is machined with a flat slot for the setscrews to contact. Thread-locker is used on all high stress parts. When assembled, the components are locked and resist twisting from the high torque servos, and the servo axes are precisely located. The lens base-plate is also slotted to support the focus servo to ensure the focus adjustment works consistently. The focus servo travels with the lens base-plate and operates regardless of tilt.

An ideal design places the servo axes equal distances from the image center and at equal separation angles so that the servo torques are equal, and the servo movements required to adjust a given tilt are the same. The filter wheel used in the Evryscope is a rotating carousel style, with its center axis necessarily offset from the image center. This complicated the Robotilter design by requiring the servo locations to be offset from the image center in order to fit on the filter wheel top and within the camera mounts. We mitigated this constraint somewhat by orientating the servo bodies toward the filter wheel space allowing the servo rotation axes to be moved closer to the image center. The arrangement features two opposing servos and a central one with a different lever arm and torque demand. These differences are managed in our software (described in \S~\ref{section_software}).

\begin{figure}[tbp]
\centering
%apj
%\includegraphics[width=0.75\columnwidth]{Robotilter_1.pdf}
\includegraphics[width=0.75\columnwidth]{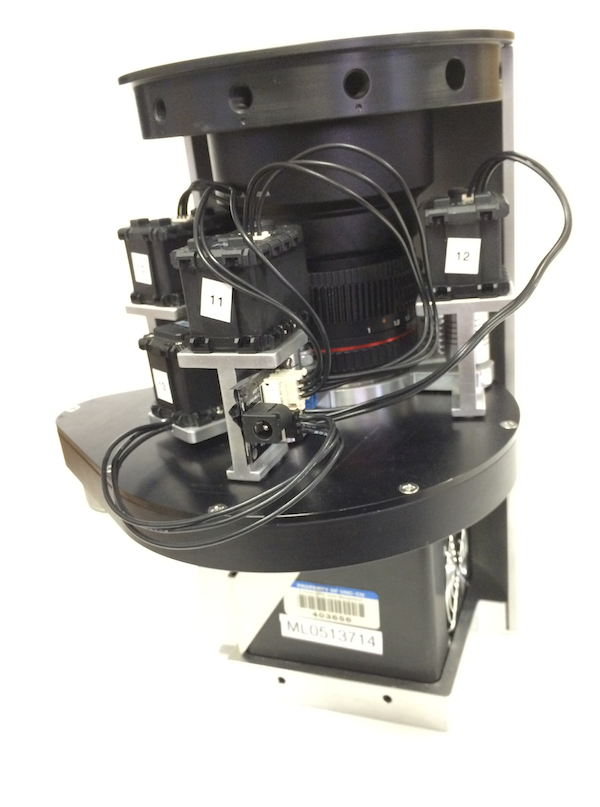}
%SPIE
%\includegraphics[width=3.5in]{Robotilter_1.png}
\caption{The Robotilter mounted in the camera mounts, fitting within the footprint of the filter wheel.}
\label{fig:robotilter_image}
\end{figure}

The Robotilter assembly mounts to the top plate of the filter wheel to avoid costly re-configuring of the existing filter wheel, CCD, or camera mounts. The footprint of the mechanism is contained within the camera mounts (Figure \ref{fig:robotilter_image}) so that the tight packing of the cameras in the mushroom is unchanged. The Robotilter upgrade was completed entirely on mountain and with minimal down time.

\subsection{Electrical Design}

Each Robotilter is powered by a 12V input line which is supplied by the accessory power supply units on the panels located in the sides of the mushroom. Each of the servos is connected to the input power line and are operated sequentially to limit total current drawn. A separate signal line connects all the servos to USB control boards which in turn are routed to the control computer. Multiple Robotilters form a serial-addressed network containing up to 14 cameras (56 servos) to reduce the number of wires, boards, and USB cables routed to the control computer. Communications over the serial line follow the Dynamixel half-duplex protocol; the large number of devices and line branches required a reduction in baud rate to 9.6kbps to enable error-free transmission (this speed is not a limiting factor for the system operation).

%----------------------------------------------------------------------------------------
%	SOFTWARE SOLUTION
%----------------------------------------------------------------------------------------

\section{The Robotilter Software Solution} \label{section_software}

In order to find the optimal image quality, the Robotilters need to position the lens to the theoretical starting point and explore in 4 dimensions (x and y tilt, separation, lens focus) to find the optimal combination. Image quality must be expressed in mathematical terms while consistently capturing the tilt, focus, focal plane, and PSF aberrations in order for a software tilt and focus solution to work properly.

\subsection{Potential Approaches} \label{subsection_conventional_approach}

Several approaches could seemingly remove image tilt and optimize focus. We discuss the most obvious ones (to us) here, our first attempts to solve the problem, and how the pixel scale and PSF distortions from the very wide field influenced our final solution. It is reasonable to begin with a conventional approach, but the Robotilters are a unique instrument and it is difficult to define a "conventional" approach to guide the software solution. However, we can combine conventional elements of image quality measurement and optics alignment as a beginning point. We first discuss this method, variations tried, and the insights learned that helped develop the final solution.

\subsubsection{Conventional Approach}

The most conventional approach we can imagine would use standard PSF quality measurements (FWHM and Strehl) to characterize image quality per region to test the parameter space and converge on a solution that maximizes total image quality. Although the Strehl is typically used for diffraction-limited images, very far from the Evryscope image quality, we use the definition here in an analogous way. Strehl in the Evryscope images measures the amount of light contained the peak of the PSF, and is thus a measure of encircled energy in the region of the PSF most important for determining the system’s limiting magnitude.

There are two primary challenges with this conventional method. First, how to test the parameter space, and second how to measure \textit{total} image quality. Using a simple grid-parameter search is ideal, however, it was not obvious how to define a grid that would test all the parameters (two tilt axes, lens / CCD separation, and lens focus position) without being overly complicated or vulnerable to degeneracies. Using a random parameter search, such as an exploratory simplex algorithm, could in principle deal with these issues but with the challenge of avoiding local minimums. Measuring total image quality is also challenging - as shown in \S~\ref{sub_section_image_quality_maeasurements}, out of focus sources show different distortions if they are above or below the image plane, adding to the difficulty in comparing quality.

\subsubsection{Original Approach}

The original strategy we tried was a variation of the approach described in the preceding paragraph. We used the FWHM and Strehl to measure image quality in several regions and compute a total image quality, combined with a simplex algorithm to explore different tilt, lens / CCD separation, and lens focus position to maximize total image quality. This approach failed for several reasons - the difficulty in capturing PSF quality, defining total image quality, and avoiding local minimums. We found the FWHM and Strehl ineffective at reliably displaying image quality in the presence of tilt and with coarsely sampled PSFs. In many cases a particular region can be optimized at the expense of another and return a higher total image quality measurement, adding to the difficulty in identifying images with significant tilt. The exploratory approach frequently converged on a local minimum with less than desired results, and the solutions were rarely repeatable.

\subsubsection{Modified Approaches}

We tried modifying the test space and quality metric of the alignment algorithm, however, the poor solution results and inconsistency persisted. We developed an auto-correlation quality metric that showed initial promise, but it still struggled with similar issues as the FWHM and Strehl. A grid with different tilt axes replaced the simplex algorithm to test the parameter space in a non-random way, again with similar poor solutions. Further testing revealed the challenges were independent of the camera or observing field. We then reduced the parameter space by holding the lens focus and lens / CCD separation constant, and only tested the two tilt axes. When this change did not significantly improve results, we tried visually locating the tilt axis and only exploring tilt about this axis, reducing the number of parameters to one. This modification still did not produce the desired results, and it became apparent that changing the tilt (in order to adequately test the parameter space for the best tilt) was causing a deeper dependence to emerge and prevent a converging solution. This dependence is best understood by analyzing the quality metric.

\subsection{The PSF Problem} \label{susection_psf_problem}

In order to measure PSF quality, the FWHM and Strehl measurements require a finely sampled PSF and the FWHM assumes a symmetric profile. Wide-field images with potentially severe field aberrations (including coma and astigmatism) and coarse pixel sampling do not have these characteristics. Images with tilt worsen the asymmetry and significantly reduce the effectiveness of the FWHM and Strehl to reliably displaying image quality in this situation. The main challenge is the structure of the PSF core and halo, and the coupled effect that the focus, tilt, signal, and CCD position have on each of them. 

\begin{figure*}[tbp]
\centering
\includegraphics[width=2.0\columnwidth]{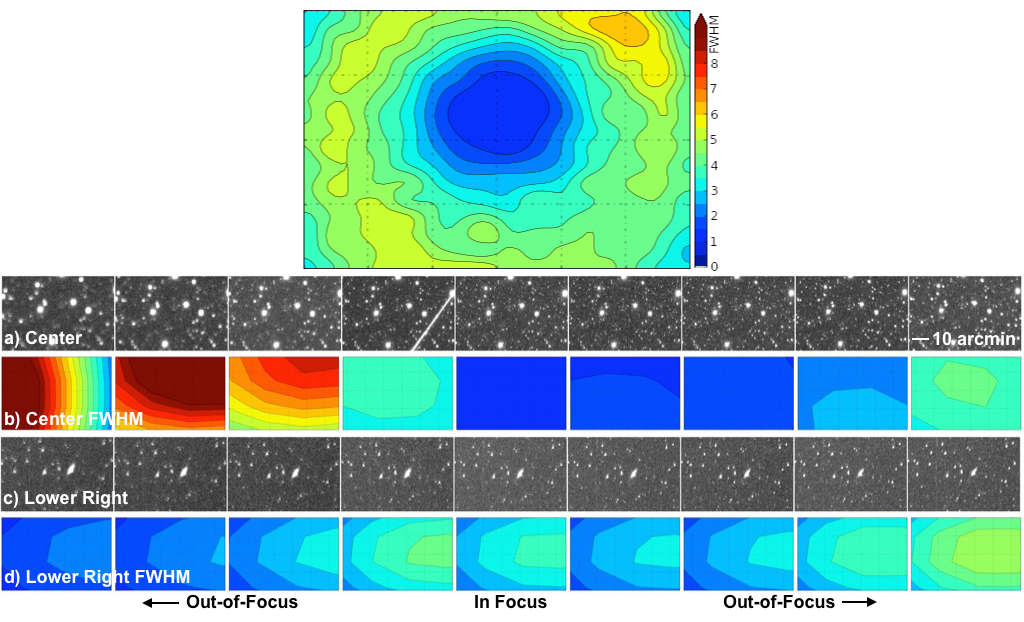}
%SPIE
%\includegraphics[width=1.0\columnwidth]{robo_sweep_combined_final.png}
\caption{\textit{Top:} The FWHM plot from an Evryscope wide-field image with little tilt. \textit{Bottom:} A sweep of images with different focus positions, the center columns are in focus while the left columns are out-of-focus below the focal plane and the right columns are out-of-focus above the focal plane. The steps between images is constant at several times larger than the sub 10 $\mu m$ level necessary to remove tilt (chosen to aid in visualization). \textit{a)} 300 x 200 pixel closeups of the center region of the images. \textit{b)} The FWHM of the same center region. Three focus positions show similarly good quality, and the response is different below and above the focal plane. \textit{c)} 300 x 200 pixel closeups of the lower right corner region of the images. \textit{d)} The FWHM of the same lower right corner region. The quality metric struggles to discriminate between the focus positions, finds more than one minimum, and the best quality is located at the very out-of-focus position shown on the far left. These issues are exaggerated in images with tilt. The challenge in capturing quality in wide-field, large pixel images with tilt led us to develop a custom tilt driven quality metric that analyzes the images as a grid and uses a predetermined movement sequence to capture images for analysis.}
\label{fig:quality_combined}
\end{figure*}

Consider an average PSF in the center region of a well focused wide-field, coarse pixel image, with little tilt as shown in Figure \ref{fig:quality_combined}. The PSFs (row a, middle column) are narrow with almost all of the flux contained within the central pixel, and with a symmetric halo that is insignificant for most sources. Changing the focus (for this discussion by adjusting the lens / CCD separation) results in a widening of the PSF to form a blob that is again symmetric and still limited in extent, shown by the right columns in row a. Changing the focus in the opposite direction gives a similarly widened PSF blob, mostly indistinguishable from the first focus movement. The separation change between columns is a constant 20 $\mu m$, considerably larger than the level the Robotilters are intended to remove, chosen for better visualization. The FWHM measured for the center region (row b) generally captures the quality change due to the focus change, but not at the level necessary as demonstrated by the very similar measurements in columns 5-7. The focus sweep, also known as a through focus sequence, demonstrates a further challenge of the FWHM measurement - the FWHM does not change significantly around the focused position. This is to be expected in situations with minimal aberrations (as exist in the image center) and coarse sampling. The FWHM response curve from the focus sweep is parabolic, and is in the shallow region of the parabola when the image is in focus, with little discrimination between changes in this narrow parameter space. Testing a wide focus range is necessary to estimate the best position. The fit is also asymmetric as shown by the difference in response below and above the focal plane (shown in the plot as steep on the left and gradual on the right columns).

The situation is more challenging for sources located in regions other than the center of the image. Consider an average PSF on the lower right corner of the same camera displayed in row c of Figure \ref{fig:quality_combined}. The center columns show elongated and distorted PSFs, even thought the region is in focus. A significant amount of the flux is in the halo. The PSFs in the right columns are out of focus but also elongated and distorted in a more severe way. Most of the flux is contained in the halo, which is not symmetric. The images on the opposite side of the focal plane are affected differently, as shown in the left columns. Here a significant fraction of the flux is in the core with a dispersed halo in addition to being distorted. As the lens is moved further, the signal decreases so severely that the halo disappears and only a faint core is detectable. The number of sources also decreases significantly.

Unfortunately, from a quality standpoint the PSF appears narrow and with almost all of the signal in the center pixel. There is very little discernment in image quality across the focus positions (which spans $\approx$20 times the level of tilt the Robotilters are designed to remove). More troublesome, the position of best measured quality is the far left column corresponding to a very out-of-focus position, driven by the disappearing halos and dim sources. This position of a severely unfocused region scores high from a traditional quality metric. The FWHM measurements are shown, but we found the Strehl suffers from similar issues with different best positions.

It is also important to point out that the best and worst quality of each region (the center versus the lower right corner as shown in Figure \ref{fig:quality_combined}) are in no way comparable. The best quality in the center region is much better than the best in the lower right corner, with a similar disparity in worst qualities. This behavior is expected in a system with field aberrations (including coma and astigmatism), and are worsened with the fast Evryscope optics. Here we have shown the center and lower right corner regions, other regions suffer from similar challenges and manifest in different ways.

If we adjust the image tilt (while holding the image center fixed) the lower right corner PSFs will distort differently than before due to the changed tilt, but the focus will also change. This can be seen in the lower right corner shown in Figure \ref{fig:image_quality_field} of the same camera but with significant tilt. Now the perceived PSF quality will depend on the coupled tilt and focus effects, the loss of signal, and the compromised halo. Adding to the difficulty, the tilt and focus challenges vary by region. In this example camera, the PSF shape in the upper left corner is approximately opposite of that in the lower right corner, and the PSFs in the edges are completely different than those in the corners or center.

Moving image tilt to explore the parameter space changes the PSF characteristics, is coupled to the focus, and the PSF effect is region dependent. This places a severe burden on a quality metric. Even without changing image tilt for comparison, measuring quality in a tilted image is challenging as is comparing quality across regions.

To summarize the ideas in this section, standard techniques (described above as well as additional commonly used metrics that were tested but not discussed) failed to capture PSF quality. With the Evryscope lens and CCD package the PSF halo tends to degrade so rapidly when out of focus, that it becomes undetectable. This leaves only the small core that appears as good image quality but actually only encloses a small amount of the signal. Only limited sources in the Evryscope images are bright enough to counter this challenge and the common methods optimize the size of the small cores, which drives the image further out of focus.

\subsection{The Robotilter Approach}

From the challenges described in \S~\ref{subsection_conventional_approach} and \S~\ref{susection_psf_problem} we developed the Robotilter software solution that uses a custom tilt driven quality metric, analyzes the image as a grid, and uses a predetermined movement sequence to capture images for analysis. The solution reliably removes tilt (from normal manufacturing and installation tolerances) and optimizes image focus in the same step, is repeatable, and takes approximately 2 hours to run. Cameras with excessive initial tilt (starting values far from optimal) benefited from additional optimization runs. We found that in this situation, the algorithm iteratively converges to the the optimal solution after repeated runs. We describe the process below.

\subsubsection{Tilt Driven Quality Metric} \label{combo}

We developed a new image quality metric designed to measure quality in the presence of image tilt and to differentiate sources above and below the focal plane. The quality metric is combination of standard PSF measurements, custom PSF measurements, and regional measurements to give a combination score which we call the \textit{combo}. The \textit{combo} is calculated for a small ($\approx$1 sq. deg.) region of an image by calculating different quality measurements and multiplying the normalized values for an overall region score. The algorithm uses Source Extractor \citep{1996A&AS..117..393B} to extract data from sources in the region. To filter out the dim and poorly sampled sources, we require the detections to be greater than 4 $\sigma$ above the background, not have blending flags, and the PSFs to comprise at least 15 pixels. Here the pixels are defined to be part of the PSF if they are above the background limit, and increasing the photometric aperture (the circle used to define the pixels included in the PSF, hereafter photometric aperture) size by one pixel does not increase the number of pixels in the PSF.

For each of the filtered sources, we calculate the PSF FWHM and Modified Strehl ($M_S$), which is the standard Strehl scaled by a multiplication factor appropriate for the Evryscope (this is not a real Strehl but still forms a useful metric component). Although the $M_S$ still struggles with the coarse Evryscope pixel scale, the multiplication factor effectively normalizes it near a value of 1 in conditions of peak quality. As discussed below, the FWMH is also modified (inverted) and combined with other elements similarly scaled so that each component contributes similarly. We also calculate custom PSF measurements for each source. 1) The Radius Ratio ($R_R$): defined as the photometric aperture radius required to enclose all pixels in the PSF divided by the ideal radius necessary to enclose the same number of pixels in the PSF, if the PSF was perfectly round. 2) The Distortion Factor ($D_F$): defined as the average distance of the pixels in the PSF from the PSF center divided by the the ideal average distance the same number of pixels in the PSF would be from the center, if the PSF was perfectly round. We then calculate the average FWHM, $M_S$, $R_R$, and $D_F$ for each region. We also count the number of filtered sources for each region and normalize across the image sweep ($N_S$). The FWHM, $R_R$, and $D_F$ average quality elements are inverted so that all factors treat a higher number as a higher quality. They are combined to give the \textit{combo} quality for the region:

\begin{equation}
\textit{combo} = \left(\frac{1}{FWHM}\right)M_S\left(\frac{1}{R_R}\frac{1}{D_F}\right)N_S
\end{equation}

The \textit{combo} metric benefits from the pooled effectiveness of the different elements to offset a particular ineffectiveness of an individual element. The FWHM measurement tends to capture out-of-focus PSF quality on one side the focal plane but fails on the opposite side. The Strehl tends to perform similarly but in the opposite way as the FWHM. The radius ratio, distortion factor, and number of sources tend to capture the variation in quality regardless which side of the focal plane the PSF is unfocused, but they are not discriminatory enough by themselves to capture the quality of the region. When combined as in the \textit{combo} equation, the metric effectively captures the region quality especially on tilted images. We demonstrate in \S~\ref{section_combined_solution} the \textit{combo} solution converges on all regions of the Evryscope images when used in our full solution algorithm. 

\subsubsection{Analysing the Image as a Grid} \label{grid}

Images are split into a 16 x 24 grid resulting in 384 regions with $\approx$1 sq. deg. FOV each. This grid size is chosen to obtain a fine enough sampling of the image field and to have enough bright stars in each region. For each region, the quality is calculated using the \textit{combo} (\S~\ref{combo}) metric. The \textit{combo} scores are not compared across regions, they are instead captured for each of the 384 regions of a particular image. The servos are moved and the \textit{combo} scores captured for each region of the new image. Regions can be compared across images to determine if a servo movement helped or hurt the quality of each part of the image independently.

\subsubsection{Predetermined Movement Sequence} \label{sweep}

We use a predetermined sequence (which we call the focus sweep) to move the servos and acquire a series of images for the Robotilter solution. The critical idea of this approach is to hold the tilt constant and only change the lens separation distance. Admittedly, this is counter-intuitive. A conventional approach (\S~\ref{subsection_conventional_approach}) adjusts the tilt and separation to explore those parameters and search for an optimal solution to the tilt and separation. While it seems reasonable to adjust the parameters that are to be optimized; in this case the image quality, focal plane, and local minimum challenges described in the previous sections are prohibitively difficult to overcome. The focus sweep approach avoids these problems altogether by instead finding the best focus for each region. It is similar to focusing a standard telescope - sweep in distance over the potential focus range and move to the position of best quality. In the Robotilter focus sweep we capture the best position of each region independent of the tilt since it is constant over the image.

\subsubsection{The Combined Solution} \label{section_combined_solution}

\begin{figure*}[tbp]
\centering
%apj
%\includegraphics[width=1.0\columnwidth]{ML0453714_20161006_combo2_3072_2304.png}
%\includegraphics[width=1.0\columnwidth]{ML0453714_20161006_robotilter_si_data_2.png}
\includegraphics[width=2.0\columnwidth]{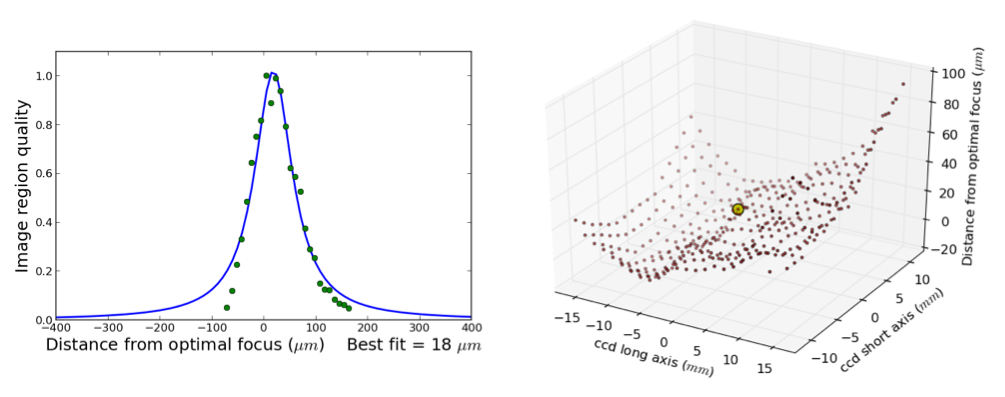}
%SPIE
%\includegraphics[width=1.0\columnwidth]{combined_robo_8_all.png}
\caption{The Robotilter solution holds the tilt constant and gathers a series of images in a focus sweep by adjusting the lens / CCD separation, splits the images into a grid, and measures the quality per region as described in \S~\ref{section_combined_solution}. \textit{Left:} The quality for a small region in the center of the image as a function of the distance from optimal focus as determined by the servo positions. As the lens / CCD separation distance sweeps from a maximum to a minimum, the image quality is low and reaches a maximum value before falling off as demonstrated by the green points. We fit a Lorentzian (the solid blue line) to measure the position of the best quality (18 $\mu m$) for this region of the image. \textit{Right:} The same small region in the center of the image is shown as the yellow circle with the 18 $\mu m$ distance from optimal focus. The image is divided into 384 regions and the image quality is calculated for each region in the same way as the example in the left panel. The pixel location of the center of each region is converted to a physical position from the image center, and the information is combined to construct the focal plane (the red points) capturing the tilt and 3 dimensional nuances.}
\label{fig:robotilter_solution}
\end{figure*}

The Robotilter solution holds the tilt constant and gathers a series of images in a focus sweep, splits the images into a grid, and measures the  quality per region as described in the previous sections. Figure \ref{fig:robotilter_solution} shows the process on a representative camera. A focus sweep (holding the tilt constant \S~\ref{sweep}) of 200, 30-second images is acquired with a separation distance of 60 servo steps (4.5 $\mu m$) between each exposure. Each image in the stack is split into a grid of 384 regions (\S~\ref{grid}) and the quality of each region is calculated using the \textit{combo} metric (\S~\ref{combo}). The servo positions are determined for each region corresponding to the optimal quality; the position in servo steps is then converted to a distance. As the lens / CCD separation distance sweeps from a maximum to a minimum, the image quality is low and reaches a maximum value before falling off, and we fit a Lorentzian to measure the best position. The Lorenzian profile was an empirical fit to the data, providing a much-improved match over a standard Gaussian or parabolic fit. The choice of a Lorenzian was motivated by the need for a more-peaked function, without physical motivation. An example from the center region of a camera with significant tilt is shown in Figure \ref{fig:robotilter_solution}. The pixel position of the chip is expressed in a distance from the chip center, and is combined with the quality information to create a 3-D contour of the focal plane.

\begin{figure*}[tbp]
\centering
\includegraphics[width=2.0\columnwidth]{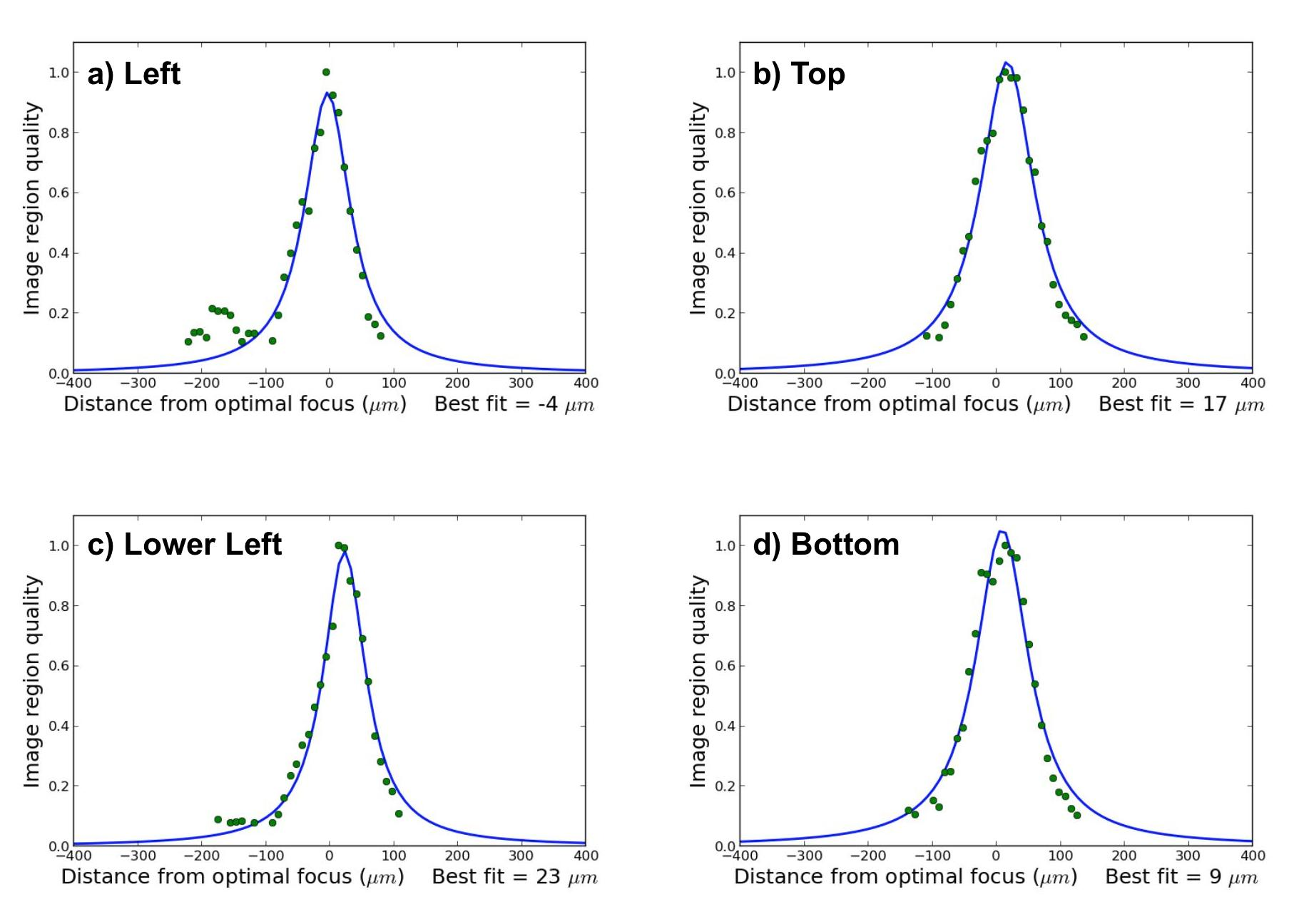}
\caption{\textit{Top: a)} The quality for a small region near the left edge of the image. The region is challenging with distorted PSFs, as illustrated by the scatter in the points and the secondary maximum near -200 $\mu m$. The feature is caused by the FWHM component struggling to accurately measure quality in this circumstance (out of focus below the focal plane). The robustness of the \textit{combo} quality metric is demonstrated by the ability to overcome the shortcomings of a single element by pooling all of the elements, and by scaling the elements so that one does not dominate. The best fit is accurate for the region and is consistent with the best fit in nearby regions and with the overall focal plane. \textit{b)} The quality of the top of the image. \textit{c)} The quality of the lower left corner of the image. \textit{d)} The quality of the bottom of the image.}
\label{fig:robotilter_solution_2}
\end{figure*}

Additional examples of the \textit{combo} quality metric are shown in Figure \ref{fig:robotilter_solution_2} for the top, bottom, edge, and corner of the image. The quality measurements converge regardless of region, despite the challenges from the tilted images, regional differences, and inconsistent PSFs. The image sweep provides 200 data points with fine separation, which aids in the accuracy of the quality fits. Additionally, points at far distances from optimal focus provide such low quality that they help constrain the base of the Lorentzian fit. These very low quality points (not shown in the plots) are flagged using a low source and high FWHM threshold, and are assigned a low value near zero. We found this to be an effective way to aid in the automated Lorentzian fit and to focus the peak width. We experimented with Gaussian and parabolic fits, but found them less reliable, and more prone to wider peaks with less accurate results.

\begin{figure*}[tbp]
\centering
%apj
%\includegraphics[width=1.0\columnwidth]{ML0453714_20161006_robotilter_si_data.png}
%\includegraphics[width=1.0\columnwidth]{ML0453714_20161006_robotilter_si_plane_data.png}
%\includegraphics[width=1.0\columnwidth]{ML0453714_20161112_robotilter_si_plane_data.png}
%\includegraphics[width=1.0\columnwidth]{ML0453714_20161112_robotilter_plot_surface.png}
%SPIE
\includegraphics[width=2.0\columnwidth]{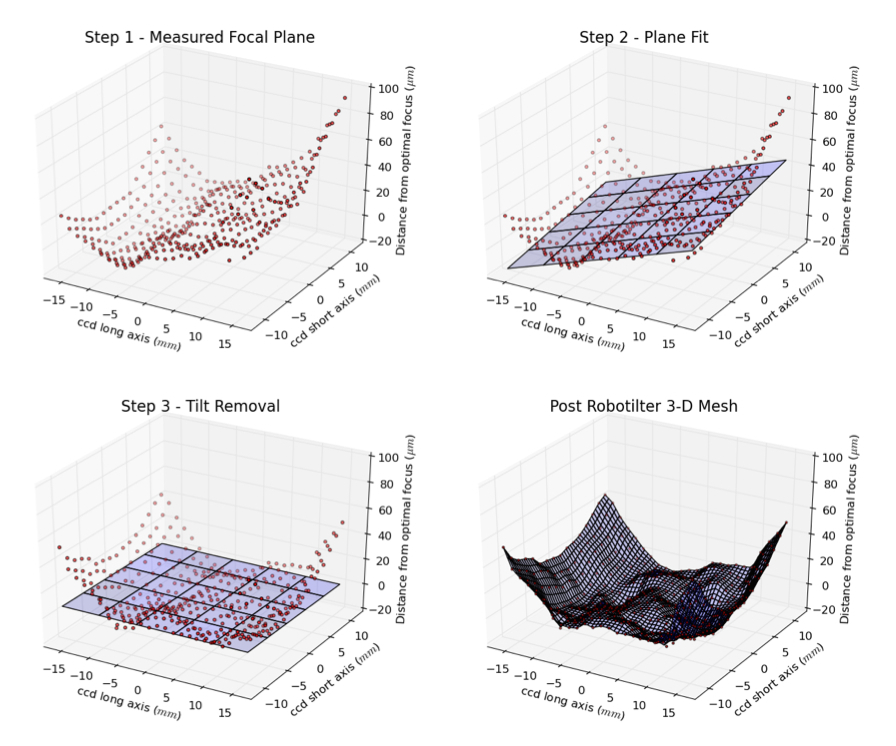}
\caption{\textit{Step 1:} The measured 3-D contour focal plane as described in \S~\ref{section_combined_solution} and Figure \ref{fig:robotilter_solution}. \textit{Step 2:} The plane fit to the measured 3-D contour focal plane. \textit{Step 3:} We move the servos so the fit plane is co-planar to the xy-plane. In this way, the tilt between the lens (fit plane) and CCD (xy-plane) is removed. \textit{Bottom Right:} The detailed mesh plot of the measured 3-D contour focal plane taken after the Robotilter solution for the same camera.}
\label{fig:robotilter_solution_plane_fit}
\end{figure*}

To remove the tilt a plane is fit (shown in blue) to the measured 3-D contour focal plane using the \textit{Scipy} module, shown in Figure \ref{fig:robotilter_solution_plane_fit}. Using the locations of the Robotilter servo axes relative to the center of the CCD (from the Robotilter mechanical design), we calculate the distance of the fit plane at each servo axes from zero (z=0). We move the servos by the calculated amounts but in the opposite direction. This moves the fit plane so that it is co-planar to the xy-plane. In this way, the tilt between the lens (fit plane) and CCD (xy-plane) is removed. An image sweep taken after the Robotilter solution producing the untilted 3-D contour for the same camera and field is shown in Figure \ref{fig:robotilter_solution_plane_fit}. In most cameras and fields, we are able to measure focal plane features and remove tilt at the sub 10 $\mu m$ level (as measured from opposite edges of the Robotilter servo shafts). We show in \S~\ref{section_combined_results} that this level of correction removes PSF differences in opposing corners and edges to the level necessary to avoid large photometric apertures (with similar size apertures needed for opposing corners and edges), and to increase the limiting magnitude by .5-1 magnitude depending on the region and amount of tilt.

We experimented with fitting more complicated shapes (paraboloids for instance), but found they did not capture the tilt in a more robust way than the simple plane fit, and they were more prone to fail in a catastrophic way. Using the plane fit offers another significant advantage - it averages the best focus across the image. Because the fit plane slices the measured 3-D contour by minimizing residuals, it finds the best overall image focus instead of maximizing one region at the expense of the rest of the image. Thus with the plane fit approach, we remove image tilt and simultaneously optimize the focus of the image field.

\subsubsection{Focal Plane}

Camera lenses offer a wide range of focus settings, the Rokinon lenses used on the Evryscope can focus from 1 meter to infinity. The lens focus mechanism (turning the lens body relative to the lens base) can actually go slightly past the infinity mark, common in photographic lenses as a margin to cover the infinity focus in the event of temperature changes. The focus servo used on the Robotilters has a fine enough control that this small range in lens adjustment corresponds to $\approx$100 servo steps. We tested this range on several cameras by removing the tilt and optimizing the focus with the lens focus at slightly different positions. In this way, we test the flatness of the field (unrelated to tilt and only dependant on lens focus position and lens / CCD separation).

The lens focus mechanism moves a group of the lens optical elements relative to other elements. This motion is different than simply moving the entire lens relative to the CCD, as we can do by moving the three Robotilter servos. With the lens focus mechanism, we actually change the optical properties - very slightly. The focal plane position is changed, as is the focal length. The optical aberrations (chromatic, spherical, field coma, and field astigmatism) are also changed. Most relevant to the Evryscope images, the focal plane position and off-axis aberrations (coma and astigmatism) are changed. The difference in focal plane position can be compensated for by the Robotilter adjusters. Different combinations of the lens focus position and the CCD / lens separation distance (within a small in-focus range) return different image quality across the field. A focus and separation combination that results in a high quality image with a flat field is advantageous for wide-field surveys like the Evryscope since the PSFs will be closer to in-focus regardless of position on the CCD. 

\begin{figure*}[tbp]
\centering
%apj
%\includegraphics[width=0.45\columnwidth]{Close_focus.JPG}
%\includegraphics[width=0.45\columnwidth]{Far_focus.JPG}
%\includegraphics[width=1.0\columnwidth]{ML1071914_slope_all_lens_norm_gauss_color.png}
%\includegraphics[width=1.0\columnwidth]{ML0453714_20161112_focal_plane_lens.png}
%SPIE
\includegraphics[width=2.0\columnwidth]{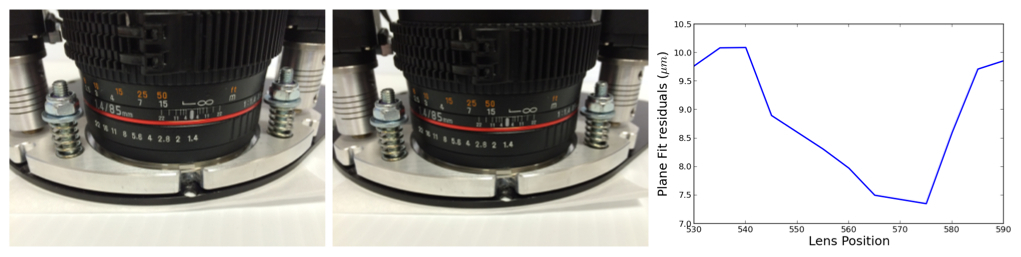}
\caption{\textit{Left:} The potential focus range of the lenses. \textit{Right:} The field flatness as a function of lens focus position (530-590 in servo position for this lens), by computing the residuals of the plane fit to the measured 3-D focal plane contour. The flattest field is at $\approx$15 servo steps from the maximum lens focus, on average for the Evryscope camera assemblies.}
\label{fig:robotilter_solution_focal_plane}
\end{figure*}

We find that for the Evryscope optics, the flattest field is located not at the max lens focus but slightly "off infinity" as shown in Figure \ref{fig:robotilter_solution_focal_plane}. The test cameras all returned similar results, and we used 15 steps off maximum lens focus as our best solution for all Evryscope cameras.

\subsubsection{On-Sky Images}

The Robotilter solution uses on-sky images for all alignment and focusing. We did experiment with in-lab alignment, but found this approach challenging with undesirable results. For the in-lab alignment approach, we used a dark room with a printout of objects (lines or synthetic PSFs) for the camera to image. Having enough different regions on the printout and sufficient objects per region was cumbersome, and the tilt removal software suffered from the same quality, focus, and convergence challenges described in \S~\ref{sub_section_image_quality_maeasurements} and \S~\ref{subsection_conventional_approach}. The focus position of the in-lab setup is necessarily much shorter than the on-sky focus position. A tilt removal solution from a lab setup based on such a large focus difference does not necessarily apply to on-sky conditions. The potential benefit from in-lab alignment is to avoid alignment during telescope time, or to avoid on-mountain troubleshooting. By testing the Robotilters in lab to verify the assembly and moving the servos to the home position, we were able to realize most of the in-lab potential benefit, and use the robust on-sky Robotilter tilt removal solution to efficiently align the cameras.

%----------------------------------------------------------------------------------------
%	ALIGNMENT RESULTS
%----------------------------------------------------------------------------------------

\section{RESULTS} \label{section_combined_results}

\subsection{ALIGNMENT RESULTS FOR ALL CAMERAS} \label{section_alignment}

\begin{figure*}[tbp]
\centering
%apj
%\includegraphics[width=1.0\columnwidth]{ML0094214_20161122_robotilter_plot_surface.png}
%\includegraphics[width=1.0\columnwidth]{ML0014214_20161117_robotilter_plot_surface.png}
%\includegraphics[width=1.0\columnwidth]{ML1051914_20161124_robotilter_plot_surface.png}
%SPIE
\includegraphics[width=2.0\columnwidth]{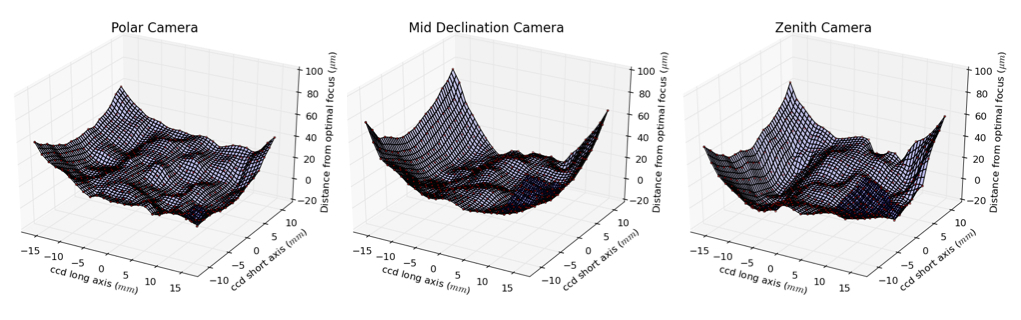}
\caption{The post-Robotilter camera alignment results for three additional cameras, distributed in declination. Shown is the polar facing camera, a mid-declination camera, and a zenith facing camera. The tilt removal is to the sub 10 $\mu m$ level. Differences in the quality and flatness of field of the optics (unrelated to lens / CCD tilt) are clearly visible.}
\label{fig:robotilter_solution_results_example_cams}
\end{figure*}

Using our software solution described in \S~\ref{section_software}, all Evryscope cameras were aligned in mid 2016 during dark sky conditions. A few cameras that were initially very far out of alignment benefited from a second run (with a smaller range and finer steps) using the initial solution as the starting point. We show 3-D contour plots for several Robotilter corrected cameras (in addition to the one shown in \S~\ref{section_combined_solution}) in Figure \ref{fig:robotilter_solution_results_example_cams}.

\begin{figure*}[tbp]
\centering
\includegraphics[width=2.0\columnwidth]{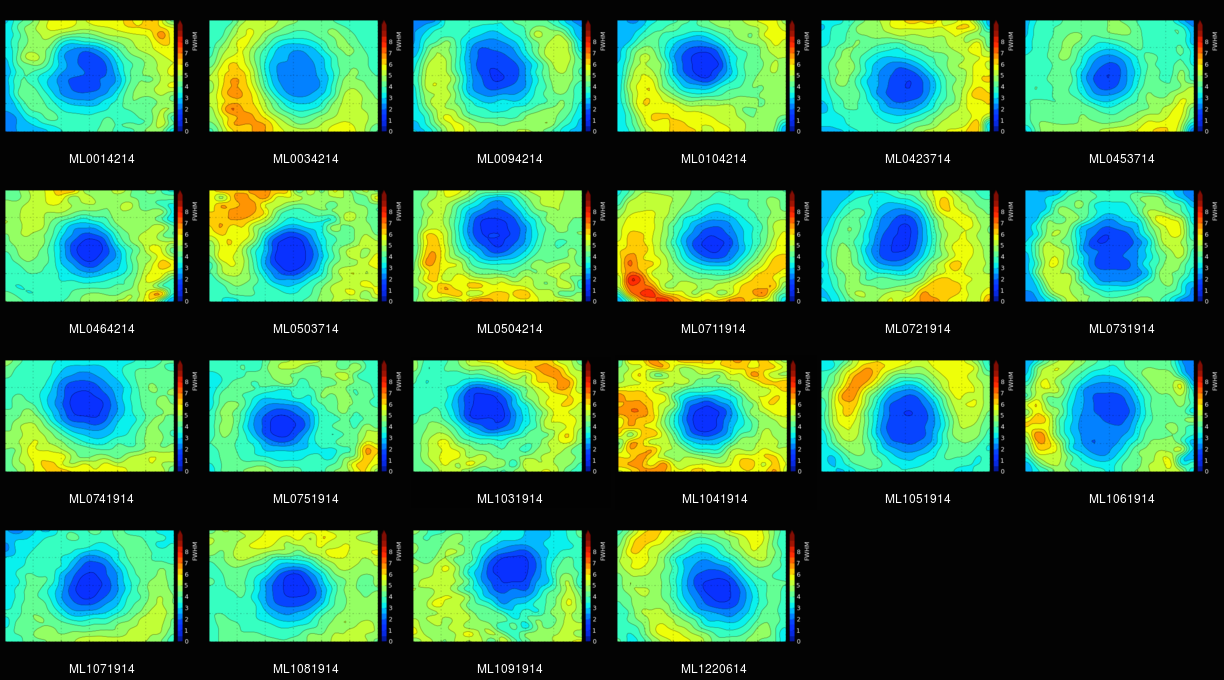}
\caption{The Robotilter camera alignment results, as shown with a daily e-mail of the FWHM display of all cameras. Although not robust enough for the full tilt removal solution, the FWHM display can be calculated on a single science image taken for each camera during the night and does not require the servos to me moved, or an image sweep to be taken. If a camera shows signs of movement, or the appearance of a very troublesome area, we can re-run the Robotilter software.}
\label{fig:robotilter_solution_results_all_cams}
\end{figure*}

The Evryscope control computer uses a scripting daemon to run the alignment algorithm. Before a nightly observation, cameras must be manually selected for alignment and placed in a queue. In order to limit power draw over 88 separate actuators, we restrict the number of camera alignments to two at a time. The post-Robotilter alignment quality for each camera was verified with the 3-D contour plot and inspection of test images taken from its Robotilter solution. Camera alignment stability is verified with a daily e-mail of a FWHM display of all cameras, a recent example is shown in Figure \ref{fig:robotilter_solution_results_all_cams}. Although limited, the FWHM display can be calculated on a single science image taken for each camera during the night and does not require the servos to be moved, or an image sweep to be taken. If a camera shows signs of movement, or the appearance of a very troublesome area, we can re-run the Robotilter software. Other than a few cameras requiring disassembly for maintenance (replacing faulty filter wheels, lenses, or cables), the aligned cameras have remained fixed since the 2016 alignment with no requirement to move even during seasonal temperature changes. 

%----------------------------------------------------------------------------------------
%	IMAGE IMPROVEMENT
%----------------------------------------------------------------------------------------

\subsection{IMAGE QUALITY IMPROVEMENT} \label{section_image}

Upon initial deployment of the Evryscope, the on-sky performance of many cameras showed a compromised image quality due to tilt and focus issues, despite careful shim-based on-sky alignment. The edges and corners of the images suffered the most, with noticeable differences in PSF shapes depending on the region and position above or below the focal plane. Here we demonstrate the improvements from the Robotilters by showing select cameras before and after the Robotilter solution.

\begin{figure*}[tbp]
\centering
%apj
%\includegraphics[width=0.49\columnwidth]{ML0094214_20150725_021554_fwhm_plot.png}
%\includegraphics[width=0.49\columnwidth]{ML0094214_20151121_070202_fwhm_plot.png}
%\includegraphics[width=0.49\columnwidth]{ML0094214_20160113_035830_fwhm_plot.png}
%\includegraphics[width=0.49\columnwidth]{ML0094214_120_20160615_231101_plot.png}
%SPIE
\includegraphics[width=2.0\columnwidth]{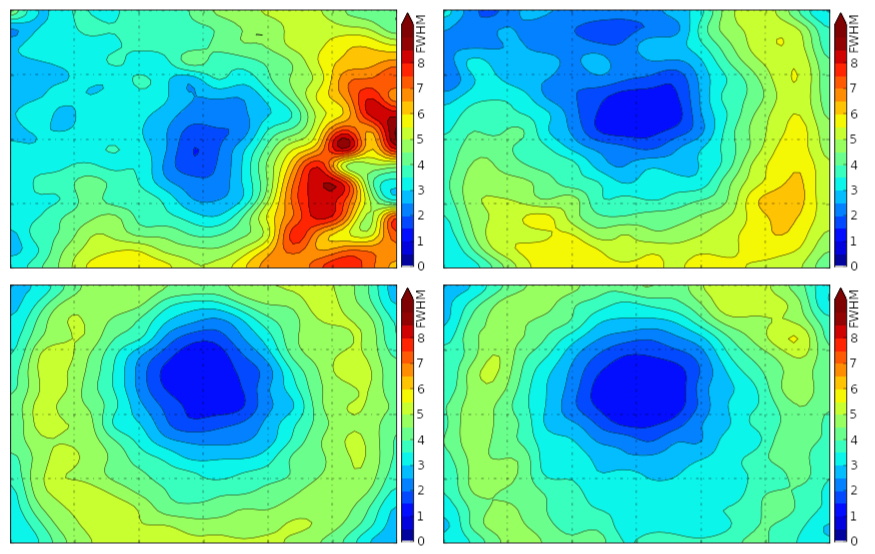}
\caption{\textit{Top Left:} Initial Deployment (pre-Robotilter) South Celestial Pole facing camera (polar camera) FWHM PSF plot. \textit{Top Right:} Same camera post Robotilter deployment, but before running software correction sequence. \textit{Lower Left:} Same camera post Robotilter correction showing the wide-scale tilt removal. \textit{Lower Right:} Same camera after the focus optimization showing the flatter field.}
\label{fig:field_psf}
\end{figure*}

Figure \ref{fig:field_psf} top left shows the FWHM plot of the camera facing the South Celestial Pole (the polar camera) upon deployment, a tilt from the lower right to the upper left corner is visible. This is the same camera described in \S~\ref{sub_section_image_quality_maeasurements} and shown in Figure \ref{fig:image_quality_field}. The top right shows the same camera after installation of the Robotilter, but before running any software tilt correction. The bottom left shows the results after the Robotilter optimization. The Robotilter upgrade improved the Evryscope image PSF FWHM and removed the wide-scale tilt. Figure \ref{fig:field_psf} bottom right shows the focus optimization results. The image quality now meets the PSF FWHM pixel target across the image with very little tilt and acceptable widening toward the edges.

\begin{figure*}[tbp]
\centering
%apj
%\includegraphics[width=1.0\columnwidth]{combined_robo_1ws.jpg}
%SPIE
\includegraphics[width=1.5\columnwidth]{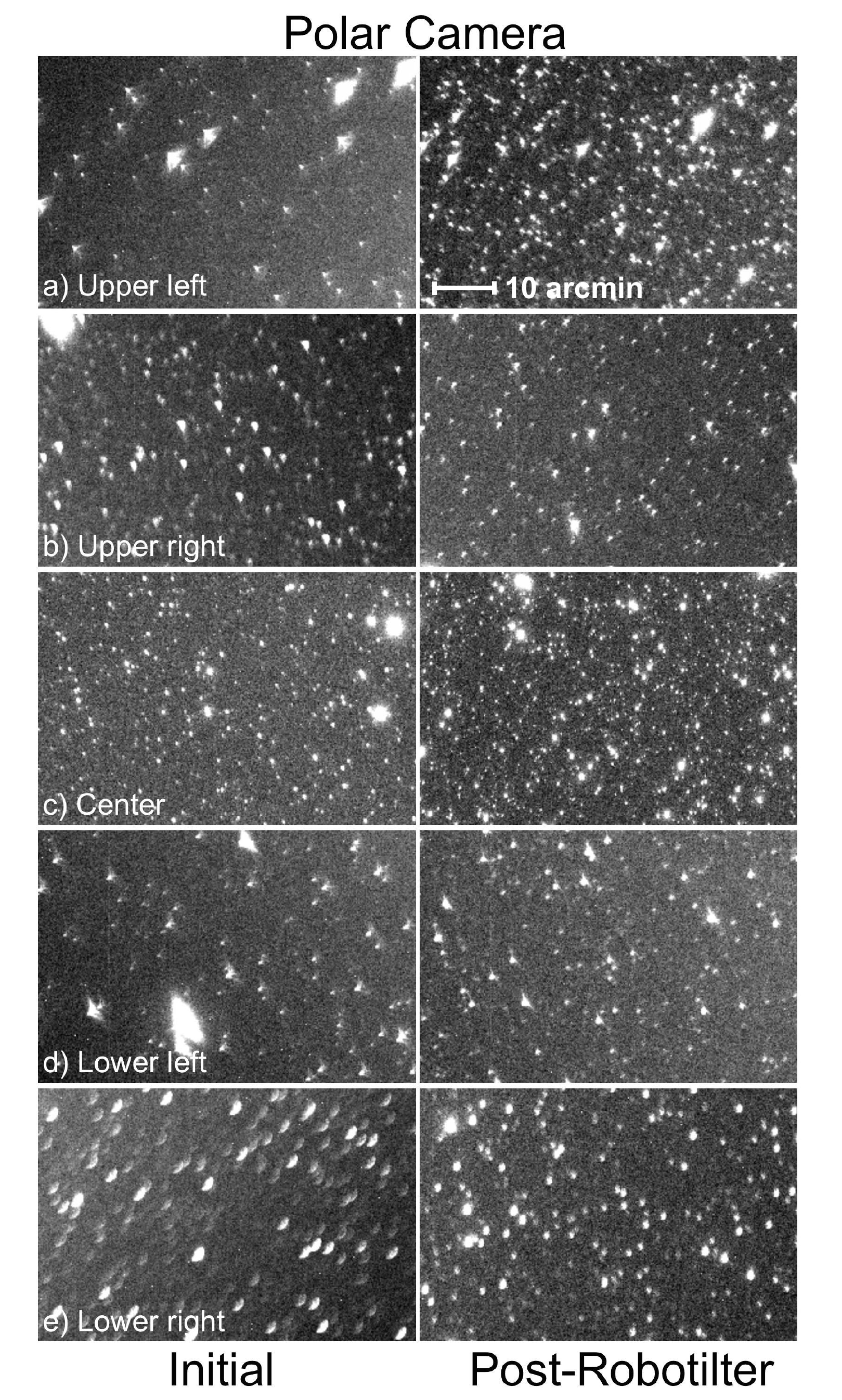}
\caption{\textit{Left:} Initial Deployment (pre-Robotilter) polar camera PSF closeup of the problematic corners. \textit{Right:} Same camera post Robotilter correction showing improvement in size, shape, and focus.}
\label{fig:combined_1}
\end{figure*}

Figure \ref{fig:combined_1} shows 300 x 200 pixel closeups of the problematic corner regions of the polar camera, before and after the Robotilter solution. The upper left and lower right corners are especially troublesome, with severe corner to corner tilt and with opposing corners on opposite sides of the focal plane. The unfocused and poorly sampled PSFs are improved by the Robotilter solution in shape and brightness. The flux is more concentrated in the PSFs, dimmer stars are visible, and more sources are detected in the images. The image improvements are realized without negatively affecting the central region.  

\begin{figure*}[tbp]
\centering
%apj
%\includegraphics[width=1.0\columnwidth]{combined_robo_3ws.jpg}
%SPIE
\includegraphics[width=1.5\columnwidth]{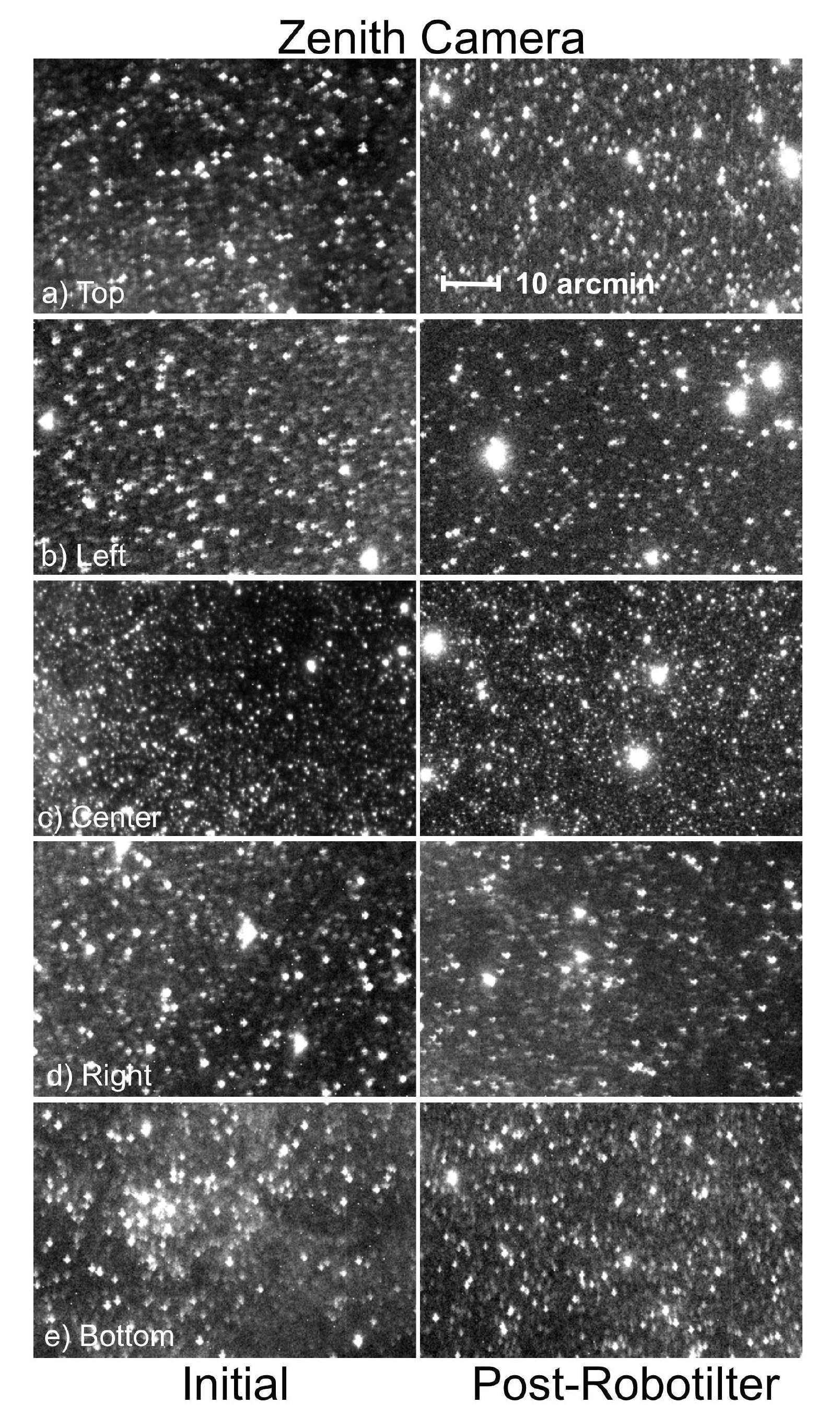}
\caption{\textit{Left:} Initial Deployment (pre-Robotilter) zenith camera PSF closeup of the problematic edges. \textit{Right:} Same camera post Robotilter correction showing improvement in quality consistency across regions.}
\label{fig:combined_3}
\end{figure*}

Figure \ref{fig:combined_3} shows 300 x 200 pixel closeups of the problematic edge regions of a zenith camera, before and after the Robotilter solution. This is the same camera discussed in \S~\ref{section_software} and shown in Figure \ref{fig:robotilter_solution_plane_fit}, now showing improved results and consistent quality across the regions.  

\begin{figure*}[tbp]
\centering
%apj
%\includegraphics[width=1.0\columnwidth]{combined_robo_2ws.jpg}
%SPIE
\includegraphics[width=1.5\columnwidth]{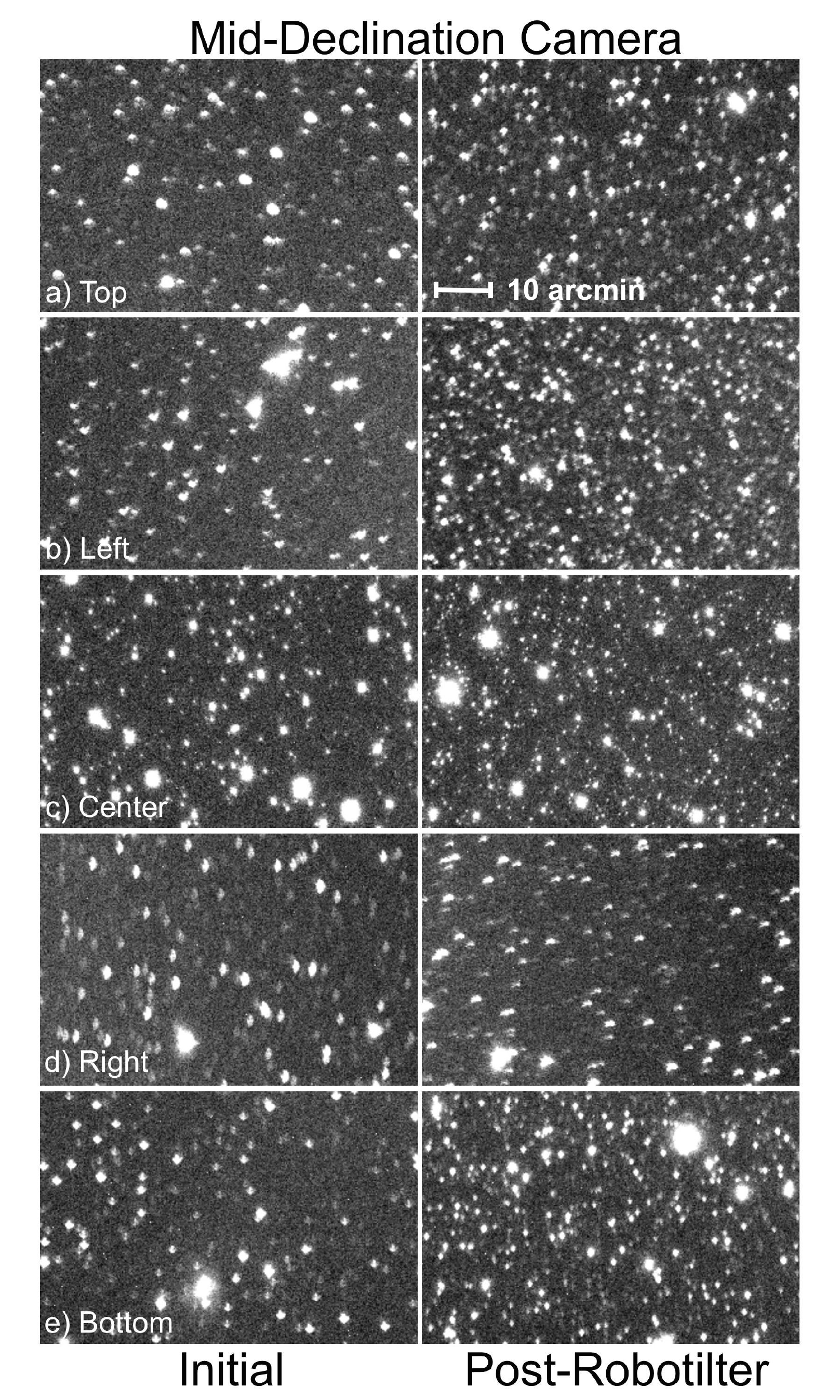}
\caption{\textit{Left:} Initial Deployment (pre-Robotilter) mid-declination camera PSF closeup of the edges. \textit{Right:} Same camera post Robotilter correction showing improvement in quality consistency across regions.}
\label{fig:combined_2}
\end{figure*}

\begin{figure*}[tbp]
\centering
%apj
%\includegraphics[width=1.0\columnwidth]{combined_robo_4ws.jpg}
%SPIE
\includegraphics[width=1.5\columnwidth]{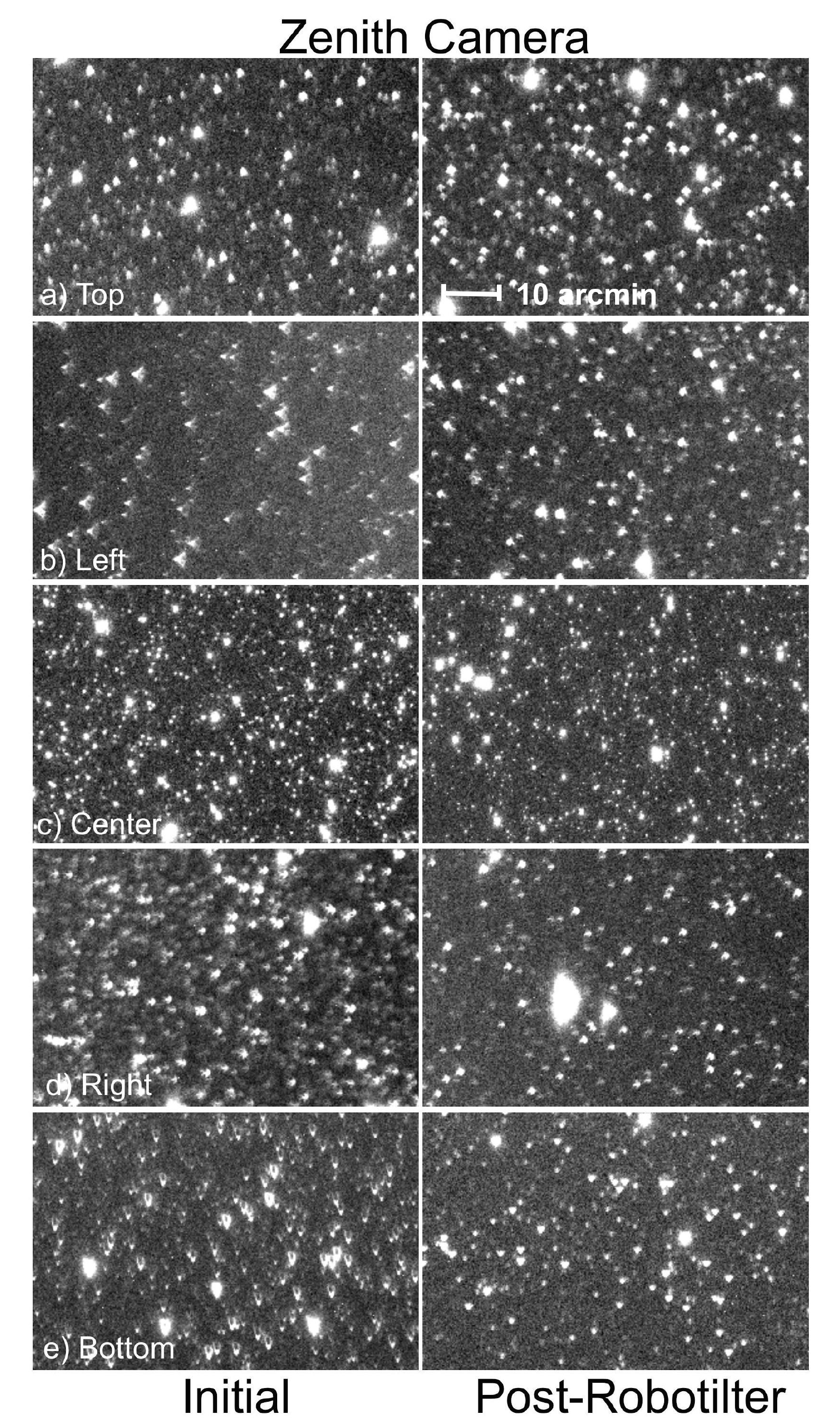}
\caption{\textit{Left:} Initial Deployment (pre-Robotilter) zenith camera PSF closeup of the edges. \textit{Right:} Same camera post Robotilter correction showing improvement in quality consistency across regions.}
\label{fig:combined_4}
\end{figure*}

Figures \ref{fig:combined_2} and \ref{fig:combined_4} show 300 x 200 pixel closeups of the edge regions of an additional zenith and mid-declination camera, before and after the Robotilter solution. These are the same cameras shown in Figure \ref{fig:robotilter_solution_results_example_cams}. The Robotilter correction again shows improvement in quality consistency across regions.

%----------------------------------------------------------------------------------------
%	EFFECTS
%----------------------------------------------------------------------------------------

\subsection{EFFECTS OF CAMERA ALIGNMENT ON EVRYSCOPE DATA} \label{section_effects}

We compared the limiting magnitude and average PSFs of images before and after the Robotilter corrections to determine the effects of the improved image quality due to the tilt removal and focus optimization. We selected cameras spread in declination (the same cameras used in \S~\ref{section_image}) and analyzed images from nights with similar dark sky, moonless, cloudless conditions. The pre-Robotilter images were collected on nights in July and September of 2015. The post-Robotilter images were taken from nights in April and July 2017. Cutouts of small regions from select images are shown in \S~\ref{section_alignment}. 

We first solve the astrometry of the images using our reduction pipeline, with APASS-DR9 \citep{2015AAS...22533616H} as our source catalog. We measure the zero point of each region in the image, and perform aperture photometry on each image to measure SNR of each source. We calculate the limiting magnitude reached by the system in dark sky conditions based on the g-band magnitude measured by APASS. The average PSF shape per region is determined using an image subtraction approach.

\begin{figure*}[tbp]
\centering
\includegraphics[width=1.6\columnwidth]{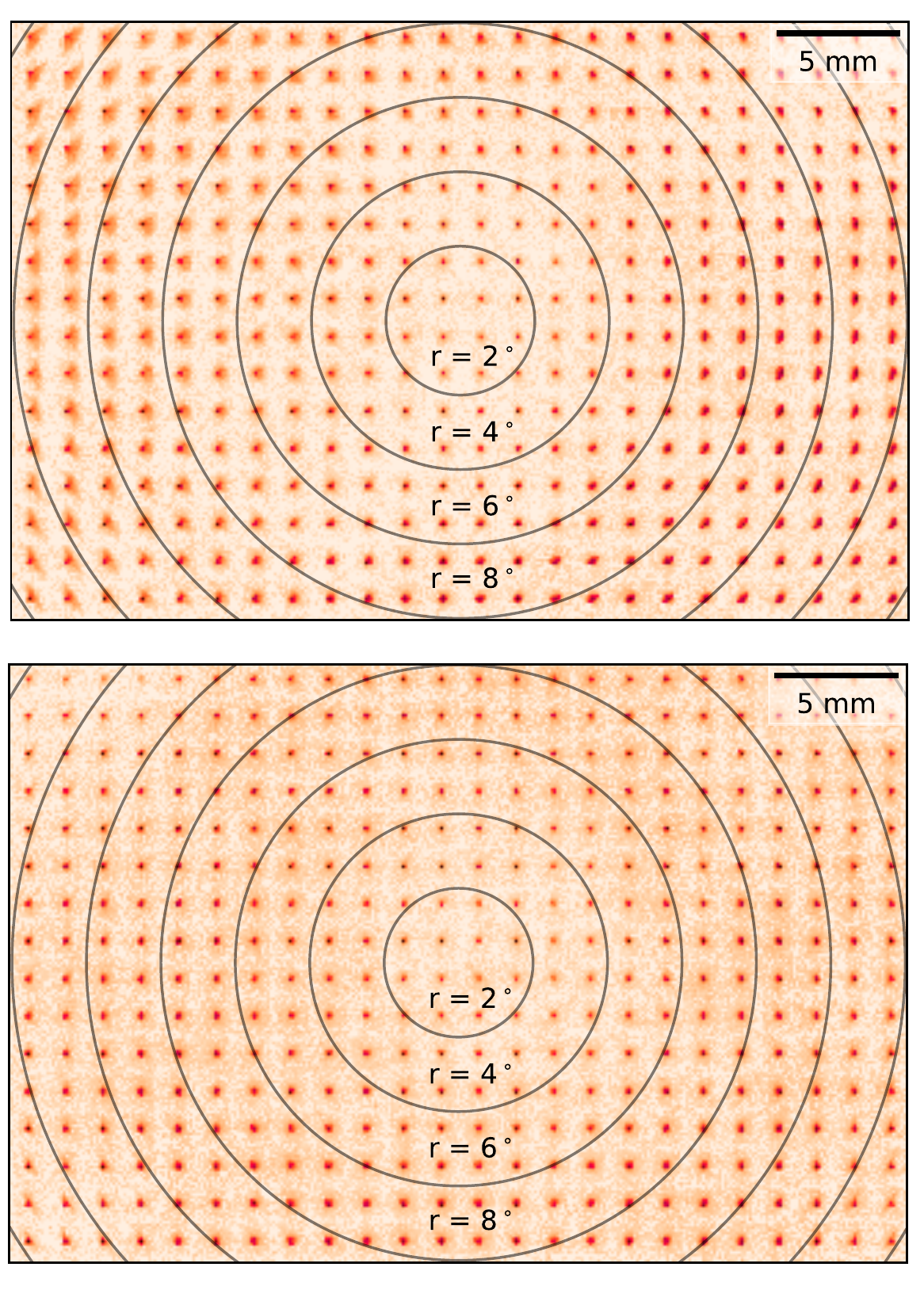}
\caption{A grid of the average PSF shape shown by region for the full field of a representative Evryscope camera. \textit{Top:} The pre-Robotilter PSF performance. \textit{Bottom:} The same camera post-Robotilter demonstrating the improved PSF consistency across the field due to the tilt removal and focus optimization. The PSF distortions are reduced, are consistent, and are symmetric about the center of the image. Compared to the pre-Robotilter image, the post-Robotilter image has an improved limiting magnitude especially on regions away from the image center. The SNR for most sources is higher (using the same photometric aperture captures a higher signal or capturing the same signal is possible with a smaller photometric aperture), and the burden on the astrometry solution is lessened by the more round PSFs (facilitating the centroiding step).}
\label{fig:psf_map}
\end{figure*}

\begin{figure*}[tbp]
\centering
\includegraphics[width=1.6\columnwidth]{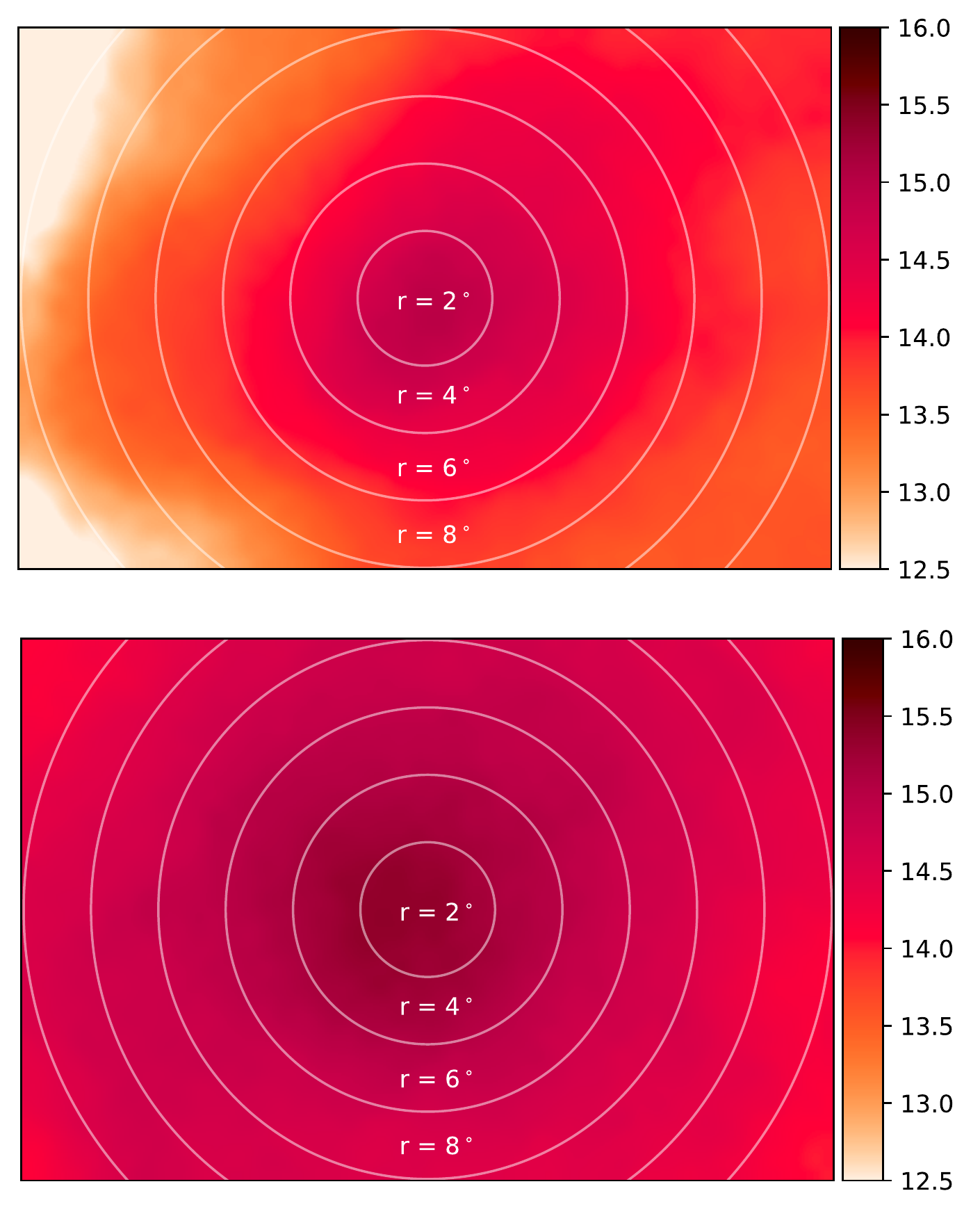}
\caption{Limiting magnitude (based on APASS-DR9 g-band) of a representative Evryscope camera. \textit{Top:} Pre-Robotilter. \textit{Bottom:} Post-Robotilter showing an improvement across the image of .5 - 1 magnitude depending on the region and the amount of initial tilt.}
\label{fig:limiting_mag}
\end{figure*}

PSF performance of a representative pre and post-Robotilter camera is shown in Figure \ref{fig:psf_map}. The PSFs from images in the corrected cameras are less distorted, especially in the corners and edges, and are smaller and more consistent across each image. The limiting magnitude improves by $\approx$ .5 magnitude in the center of the field and $\approx$ 1 magnitude in the corners. This is camera and condition dependent; we show a representative camera in dark sky conditions in Figure \ref{fig:limiting_mag}. The SNR for most sources is higher (using the same photometric aperture captures a higher signal or capturing the same signal is possible with a smaller photometric aperture), and the burden on the astrometry solution is lessened by the more round PSFs (facilitating the centroiding step).

The improved PSFs from the post Robotilter images also improves the photometric performance of the light curve pipeline; however, we did not collect sufficient pre-robotilter data to make a quantitative comparison. Additionally, other non-constant factors such as improved telescope tracking and periodic cleaning of the optics effect light curve precision and are difficult to separate.

%----------------------------------------------------------------------------------------
%	DISCUSSION
%----------------------------------------------------------------------------------------

\section{DISCUSSION} \label{section_discussion}

\subsection{Robotilter Design Improvements}

We deployed the Evryscope North (an updated version of the CTIO Evryscope) to Mount Laguna Observatory, California in November of 2018. The Robotilters installed on the Evryscope North feature several improvements. Limit switches are mounted on the lens base-plate to locate the home position in case the servos are moved out of range. A separate Raspberry-Pi single board computer controls the camera and Robotilter for each unit, allowing more than two Robotilters to be run simultaneously and reducing the number of cables and hubs. The servo piers are locked to the filterwheel top in a more robust way that locates the servo axes more precisely. The Rokinon lenses and FLI CCD cameras used on the Evryscope North are 4 years newer than those used on the CTIO system and feature mild improvements in optics and chip sensitivity. Initial image quality results from the Evryscope North point to mild improvements in image flatness and PSF quality.

\subsection{Lessons Learned}

Several lingering challenges slowed our progress over the course of the Robotilter project. The primary issues were related to assembly, servo control, and software.

The Robotilters must precisely locate and hold the 4 servos and all of the components into a small space on top of the filter wheel. This results in a considerable amount of hardware and small pieces, and the assembly is not trivial. The spring tension, shaft couplers, and the threaded shafts were the most challenging to assemble. Cycling each Robotilter assembly in the lab, for several hours over a range of servo positions, helped prevent on-sky issues. This procedure also helped identify misaligned shaft couplers or over-torqued springs. The threaded shafts needed liberal amounts of Goop lubricant to work smoothly with the brass inserts in the base-plate. Multiple cycling helped to mate the interacting surfaces, and to identify any defects that might have caused issues later. Tightening the fasteners at critical mounting points could twist the assembly resulting in the lens center not being concentric with the CCD. We made the locking slots on the servo piers more robust on the northern system which helped resist twisting. Some cameras in the CTIO system suffered from the various challenges described here, requiring on site troubleshooting. We were able to mitigate these issues in the Northern system with the minor assembly and testing corrections learned from the CTIO Evryscope. 

The servos are controllable to within $\approx$200 steps when commanded to move. The accuracy also depends on how far the servo is commanded to move and on the individual servo. This was less than ideal for the Robotilter tilt correction step, and we added a software correction to compensate for the mechanical backlash causing this servo accuracy challenge. We rely on multiple servo movements to solve this issue. In the first movement, we command the servos to move past the intended target and in a second movement to go past the target in the other direction but by a smaller amount. We then repeat this process but for a much narrower overshoot before commanding to the final position. In this way, we are able to move the servos to within 15-20 servo steps on average.

The Robotilter servos move 4096 steps per turn, and use an offset to count multiple turns. For example, one-half turn is counted as a position of 2048, and one and one-half turns is a position of 2048 plus an offset of 1. An issue that arises (and is common with servos) is in the event of a power loss the position is retained but not the offset. We addressed this issue by resetting the servo offsets to zero once the alignment was completed, so that the servo values are always within one turn, and by recording the servo positions each night.

It is possible for servos to become stuck if they are moved very far away from the home position by mistake, or if one servo is moved relative to the others that puts an extreme angle on the lens base-plate. We set the maximum servo torque low so that in the event one becomes stuck, we can manually increase the torque and move it the opposite direction to release. For the Northern system, we added independent locator switches to identify the home position and help avoid errant movements.

We underestimated the software challenge of the Robotilters, which resulted in telescope time being used for Robotilter software development. This turned out not to be a significant problem, and it was not completely avoidable. In retrospect, we might have used the pre-deployment Evryscope test camera more in the Robotilter software development. Most likely, this would have required a dedicated robotic telescope using the single Evryscope test camera. This approach would have required extra resources, and if it would have provided a benefit greater than the cost is debatable. We elected instead to deploy the Robotilters once ready, use a few select cameras to test and refine the Robotilter software, and observe with all the other cameras during that time. Once the software was completed, we aligned all the cameras and have observed continuously since then. 

\subsection{As an Optics Quality Measurement}

\begin{figure}[tbp]
\centering
%apj
%\includegraphics[width=1.0\columnwidth]{ML1091914_20161117_robotilter_plot_surface.png}
%SPIE
\includegraphics[width=1.0\columnwidth]{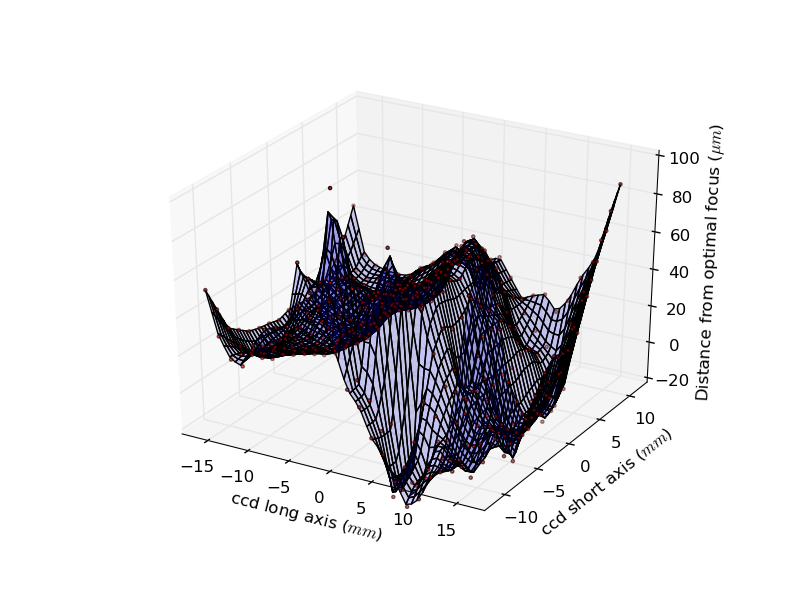}

\caption{Using the Robotilters to measure optics quality; shown is a problematic lens with an odd sheer feature visible in the measured 3-D focal plane. We replaced this lens on a maintenance trip and the camera showed an improvement in image quality. We suspect one of the lens elements was damaged (hairline crack) in transport.}
\label{fig:bad_lens}
\end{figure}

A by-product of the Robotilter solution is the precise and finely sampled 3-D focal plane, as demonstrated in Figures \ref{fig:robotilter_solution_plane_fit} and \ref{fig:robotilter_solution_results_example_cams}. The quality of the optics is clearly captured, including the flatness of the field, the image profile, regional structure, and differences between cameras. The Robotilters can be used to identify optics that will likely perform well, as well as those that could be troublesome. As an extreme example, Figure \ref{fig:bad_lens} shows a problematic lens with an odd sheer feature visible in the measured 3-D focal plane. We replaced this lens on a maintenance trip and the camera showed an improvement in image quality. We suspect one of the lens elements was damaged, possibly with a hairline crack, in transport.

\subsection{Applications for Other Instruments}

The Robotilters were designed for the Evryscope; we did not test them or simulate their potential on any other instrument. However, the Robotilter solution described in this work certainly could be adapted for use on wide field surveys using lenses or small telescopes. The basic mechanical design should scale to a variety of lens sizes and types; most likely with only simple modifications to the servo spacing, placement, and component sizes. The software solution approach using a focus sweep, image grid, and tilt driven quality metric with on-sky images should also be effective for instruments with different FOVs and pixel scales; with appropriate adjustments to the number of images, step and grid sizes, and quality metric components. It is also reasonable to consider using the Robotilter solution on larger instruments, but move the CCD instead of the optics.

\section{SUMMARY} \label{section_summary}

The Robotilter lens / CCD automated alignment upgrade was installed on the Evryscope at the end of 2015. The Robotilter hardware has performed reliably and consistently, and has demonstrated the ability to hold tilt position over several years. We developed the software necessary to align the cameras, which is specialized to remove tilt, minimize PSF distortions, optimize the focal plane, and balance focusing within the full image field. The Robotilters are completely automated, use on-sky images, remove image tilt to the sub 10 $\mu m$ level, in less than 2 hours. The tilt removal and focus optimization solutions work independent of camera or field. The Robotilter solution resulted in measurable improvements in image quality, SNR, limiting magnitude, and astrometric solutions. The average PSF extent was reduced by a factor of 2 on the edges and corners for the images, and the limiting magnitude was improved by .5 to 1 magnitude for most cameras. In this work we described in detail the challenges, development and design, software strategy, and lessons learned. 
 
This research was supported by the NSF CAREER grant AST-1555175, NSF/ATI grant AST-1407589, and the Research Corporation Scialog grants 23782 and 23822. HC is supported by the NSF GRF grant DGE-1144081. OF and DdS acknowledge support by the Spanish Ministerio de Econom\'ia y Competitividad (MINECO/FEDER, UE) under grants AYA2013-47447-C3-1-P, AYA2016-76012-C3-1-P, MDM-2014-0369 of ICCUB (Unidad de Excelencia 'Mar\'ia de Maeztu').

Special thanks to David Norris, Cliff Tysor, and Philip Thomson from the UNC machine shop for their careful work and advice.

%----------------------------------------------------------------------------------------
%	BIBLIOGRAPHY
%----------------------------------------------------------------------------------------

%\clearpage
%apj
\bibliographystyle{apj}
%SPIE
%\bibliographystyle{spiejour}
\bibliography{ratzloff_refs}

%SPIE
%\end{spacing}

\end{document}